\documentclass[epsfig,12pt]{article}
\usepackage{makeidx}
\usepackage{amsmath}
\usepackage{amsfonts}
\usepackage{amssymb}
\usepackage{graphicx}

\input epsf.sty
\makeatletter \@addtoreset{equation}{section}
\renewcommand{\theequation}{\thesection.\arabic{equation}}
\input{tcilatex}
\begin{document}

\title{\rightline{\mbox{\small {LPHE-MS-01-2021}} \vspace
{-0,2cm}} \textbf{5D} $\mathcal{N}=1$ \textbf{super QFT: symplectic quivers}}
\author{E.H Saidi\thanks{%
e.saidi@um5r.ac.ma} \ and L.B Drissi\thanks{%
\ b.drissi@um5r.ac.ma} \\
{\small 1. LPHE-MS, Science Faculty}, {\small Mohammed V University in
Rabat, Morocco}\\
{\small 2. Centre of Physics and Mathematics, CPM- Morocco}}
\maketitle

\begin{abstract}
We develop a method to build new 5D $\mathcal{N}=1$ gauge models based on
Sasaki-Einstein manifolds $Y^{p,q}.$ These models extend the standard 5D
ones having a unitary SU$\left( p\right) _{q}$ gauge symmetry based on $%
Y^{p,q}$. Particular focus is put on the building of a gauge family with
symplectic SP$\left( 2r,\mathbb{R}\right) $ symmetry. These super QFTs are
embedded in M-theory compactified on folded toric Calabi-Yau threefolds $%
\hat{X}(Y^{2r,0})$ constructed from conical $Y^{2r,0}$. By using
outer-automorphism symmetries of 5D $\mathcal{N}=1$\textbf{\ }BPS quivers
with unitary SU$\left( 2r\right) $ gauge invariance, we also construct BPS
quivers with symplectic SP$\left( 2r,\mathbb{R}\right) $ gauge symmetry.
Other related aspects are discussed.\  \  \  \  \newline
\end{abstract}


\section{Introduction}

$\mathcal{N}=1$ supersymmetric gauge theories in five space time dimensions
(super QFT$_{5}$) are non renormalizable field theories with eight
supercharges. They are admitted to have UV fixed points which can be
deformed by relevant operators such that in the infrared they flow to 5D $%
\mathcal{N}=1$ super Yang-Mills (SYM$_{5}$) coupled to hypermultiplets \cite%
{1,2}. A typical massive deformation generating this type of flow is given
by the SYM$_{5}$ term tr$\left( F_{\mu \nu }^{2}\right) $/g$_{YM}^{2}$ where
in 5D the inverse gauge coupling square 1/g$_{YM}^{2}$ has dimension of
mass. These 5D gauge theories are somehow special compared to 6D gauge
theories \cite{2aa,2ab,2ac} including maximally supersymmetric Yang-Mills
theory believed to flow to $\mathcal{N}=(2,0)$ supersymmetric 6D theory in
the UV \cite{3,1C}. \textrm{In the few last years}, super QFT$_{5}$s and
their compactification, in particular on a Kaluza-Klein circle with finite
radius and to 3D, have been subject to some interest in connection with
their critical behaviour and specific properties of their gauge phases
\textrm{\cite{1a}-\cite{7ab}}. Though a complete classification is still
lacking \cite{7b,2ac}, several examples of such gauge theories are known;
and most of them can be viewed as deformations of 5D superconformal theories
\textrm{\cite{11a,12a,13a}. }Simplest examples of SCFT$_{5}$s\textrm{\ }are
given by the so-called Seiberg family possessing a rich flavor symmetry
\textrm{\cite{14a}}; many others are obtained through embedding in string
theory. Generally speaking, this embedding can be achieved in two
interesting ways; either by using 5-brane webs in type IIB string theory
\textrm{\cite{1A}-\cite{6A}}; or by using M-theory compactification on
Calabi-Yau threefolds \textrm{\cite{1b}-\cite{5b}}. Below, we comment
briefly on these two methods while giving some references which certainly
are not the complete list since the works in this matter are\textrm{\
abundant.}\newline
The method of $\left( p,q\right) $ 5-brane webs in type IIB string theory
has led to several findings and has several features; in particular the
following: First, it gives evidence for the existence of fix point of 5D
gauge theories flowing to UV conformal points corresponding to collapsed
webs; and as such permits to study conditions for existence of critical fix
points. This web construction also indicates that not every 5D gauge theory
can flow to a SCFT$_{5}$ \cite{1}; \textrm{the existence of a SCFT
constraints the matter content of the theory}. The 5-brane method allows
also to study gauge theory dualities in 5D. This is because a given SCFT$%
_{5} $ can have several gauge theory deformations; thus generating different
(but dual) gauge theories in infrared \cite{1A}. Also, the web method
provides us with a tool to compute the instanton partition function that
captures the BPS spectrum of the 5D theory by applying the topological
vertex formalism \textrm{\cite{Nek1}-\cite{Nek6}}. It also allows the study
the global symmetry enhancements of the SCFTs \textrm{\cite{1C,2C}} and
UV-dualities \textrm{\cite{1A}}, \textrm{\cite{3C}-\cite{5C}}. More
interestingly, the 5-brane webs approach give a way to elaborate families of
5D gauge models with fix points closely related to quivers with SU gauge in
the shape of Dynkin diagrams\textrm{.} By introducing an orientifold plane
like O5-plane, the 5-brane webs can describe 5D super QFTs with flavors and
gauge groups beyond SU(N) such as SO(N) and Sp(2N) \textrm{\cite{SP1,SP2}}
as well as exceptional ones like G$_{2}$ \textrm{\cite{Nek3}}. Certain $%
\left( p,q\right) $ 5-brane webs have interpretation in terms of toric
diagrams \textrm{\cite{3A}} although, for 5D gauge theories with a large
number of flavors, they lead to non- toric Calabi-Yau geometries \textrm{%
\cite{V2}}. This brane based method is not used in this paper; it is
described here as one of two approaches to study 5D $\mathcal{N}=1$ super
QFTs underlying SCFT$_{5}$. For works using this method, we refer to rich
the literature in this matter; for instance \textrm{\cite{1A}-\cite{3A},\cite%
{1B}-\cite{6B}}. \  \  \newline
\textrm{Regarding the M-theory method, to be used in this study, one can
also list several interesting aspects }showing that it is a powerful higher
dimensional geometric approach. First of all, the 5D gauge theories are
obtained by \textrm{compactifying} \textrm{M-theory on} Calabi-Yau threefolds%
\textrm{\ (CY3) X (} resolved $\hat{X}$). Then, the effective prepotential $%
\mathcal{F}_{5D}$ and its non trivial variations $\delta ^{n}\mathcal{F}%
_{5D}/\delta \phi ^{n},$ characterising the Coulomb branch of the 5D super
QFTs, have interesting \textrm{CY3} interpretations; i.e. a geometric
meaning in the internal dimensions. The $\mathcal{F}_{5D}$ is given by the
volume $vol(\hat{X})$\textrm{\ }while its variations ---describing magnetic
string tensions amongst others--- are interpreted as volumes of p-cycles.
Moreover, the calculation of $\mathcal{F}_{5D}$ can be explicitly done for a
wide class of $\hat{X}$'s; in particular for the family of toric Calabi-Yau
threefolds like those based on the three following geometries: $\left(
\mathbf{a}\right) $ The toric del Pezzo surfaces dP$_{n}$ with n=1,2,3;
these Kahler manifolds are toric deformations of the complex projective
plane $\mathbb{P}^{2}$. $\left( \mathbf{b}\right) $ The Hirzebruch surfaces%
\emph{\ }$\mathbb{F}_{n}$ given by non trivial fibrations of a complex
projective line $\mathbb{P}^{1}$ over a base $\mathbb{P}^{1}$ \textrm{\cite%
{9b}-\cite{12b}. }$\left( \mathbf{c}\right) $ The family $\hat{X}\left(
Y^{p,q}\right) $ given by a crepant resolution of toric threefolds realised
as real metric cone on Sasaki-Einstein $Y^{p,q}$ spaces labeled by two
positive integers $\left( p,q\right) $ constrained as $p\geq q\geq 0$
\textrm{\cite{1c}-\cite{5c}}. \newline
In this investigation, we focus on the particular class of 5D supersymmetric
SU$\left( p\right) _{q}$ unitary field models based on $\hat{X}\left(
Y^{p,q}\right) $ and look for a generalisation of these quantum field models
to other gauge symmetries. Our interest into the Sasaki-Einstein (EM) based
CY3s has been motivated by yet unexplored specific properties of $Y^{p,q}$
and also by the objective of generalizing partial results obtained for the
unitary family. In this context, recall that the toric 5D super QFTs based
on $\hat{X}\left( Y^{p,q}\right) $ have unitary SU$\left( p\right) _{q}$
gauge symmetries with Chern-Simons (CS) level q. Thus, it is interesting to
seek how to generalize these unitary gauge models based on $\hat{X}\left(
Y^{p,q}\right) $ for other gauge symmetries like the orthogonal and the
symplectic. As a first step in this exploration, we show in this study that
the 5D unitary gauge theories based on $Y^{p,q}$ have discrete symmetries
that can be used to construct new gauge models. These finite groups come
from symmetries of p-cycles inside the $\hat{X}\left( Y^{p,q}\right) $. By
using specific properties of the unitary set and folding under outer-
automorphisms of p-cycles, we construct a new family of 5D SQFTs \textrm{%
having symplectic} SP$\left( 2r,\mathbb{R}\right) $ gauge invariance.
\newline
To undertake this study, it is helpful to recall some features of the
Sasaki-Einstein based CY3: $\left( \mathbf{i}\right) $ They are toric and
they extend the $\hat{X}\left( dP_{1}\right) $ and the $\hat{X}\left(
\mathbb{F}_{0}\right) .$ These geometries appear as two leading members in
the $\hat{X}\left( Y^{p,q}\right) $ family. $\left( \mathbf{ii}\right) $
They have been used in the past in the engineering of 4D supersymmetric
quiver gauge theories \textrm{\cite{1d}-\cite{4d}; and have }been recently
considered in models building of \emph{unitary }5D $\mathcal{N}=1$ super
CFTs \textrm{\cite{1e}-\cite{5e}}. $\left( \mathbf{iii}\right) $ Being
toric, the threefolds $\hat{X}\left( Y^{p,q}\right) $s and the \emph{unitary}
5D super QFTs based on them can be respectively represented by toric
diagrams $\Delta _{\hat{X}\left( Y^{p,q}\right) }$ and by BPS quivers $Q_{%
\hat{X}\left( Y^{p,q}\right) }$ describing the BPS particle states of the
unitary supersymmetric theory. \newline
The toric $\Delta _{\hat{X}\left( Y^{p,q}\right) }$ and the BPS $Q_{\hat{X}%
\left( Y^{p,q}\right) }$ are particularly interesting because they play a
central role in our construction; as such, we think it is useful to comment
on them here. We split the properties of these objects in two types: general
and specific. The general properties, which will be understood in this
investigation, are as in the geometric engineering of 4D super QFTs \textrm{%
\cite{13h}-\cite{17h}}. They also concern aspects of the Sasaki-Einstein
manifolds and the brane tiling algorithms (a.k.a dimer model) \textrm{\cite%
{1h}-\cite{10h}. }Some useful general aspects for this study are reported in
the appendices A, B, C\textrm{. The specific properties }$\Delta _{\hat{X}%
\left( Y^{p,q}\right) }$ and $Q_{\hat{X}\left( Y^{p,q}\right) }$ regard
their outer-automorphisms and the implementation of the Calabi-Yau condition
of $\hat{X}$ as well as a previously unknown property of $\hat{X}\left(
Y^{p,q}\right) $ that we describe for the leading members $p=2,3,4.$ By
trying to exhibit manifestly the Calabi-Yau condition on the toric diagram $%
\Delta _{\hat{X}\left( Y^{p,q}\right) }$, we end up with the need to
introduce a new graph representing $\hat{X}\left( Y^{p,q}\right) $. This new
graph is denoted like $\mathcal{G}_{\hat{X}\left( Y^{p,q}\right) }^{G}$ with
G referring either to the gauge symmetry SU$\left( p\right) $ or to SP$%
\left( 2r,\mathbb{R}\right) $. The construction of $\mathcal{G}_{\hat{X}%
\left( Y^{p,q}\right) }^{G}$ will be studied with details in this paper; to
fix ideas, see eq(\ref{fr}) and the Figure \textbf{\ref{y4-geo}}, the Figure%
\textbf{\  \ref{F0-geo} }and the Figure\textbf{\  \ref{y30}}.\  \  \newline
In the present paper, we contribute to the study of 5D $\mathcal{N}=1$ super
QFT models based on conical Sasaki-Einstein manifolds and their
compactification on a circle with finite radius. Using the above mentioned
discrete symmetries, we develop a method to build new 5D $\mathcal{N}=1$
Kaluza-Klein quiver gauge models based on Sasaki-Einstein manifolds $Y^{p,q}$%
. For that, we first revisit properties of the internal $\hat{X}\left(
Y^{p,q}\right) $ geometries which are known to host gauge models with SU$%
\left( p\right) _{q}$ gauge symmetry. Then, we show that some of these
Sasaki-Einstein based threefolds have non trivial discrete symmetries that
exchange p-cycles in $\hat{X}\left( Y^{p,q}\right) $ and which we construct
explicitly. By using these finite symmetries and cycle- folding ideas, we
build a new set of 5D supersymmetric gauge models based on $\hat{X}\left(
Y^{p,q}\right) $ having symplectic SP$\left( 2r,\mathbb{R}\right) $ gauge
invariance; thus extending the set of unitary gauge models for this family
of CY3. We also derive the associated BPS quivers encoding the data on the
BPS states of the symplectic theory. We \textrm{moreover show }that the
cycle- folding by outer-automorphisms generate super QFT models having no
standard interpretation in terms of gauge phases. For a pedagogical reason,
we mainly focus on the leading members of the symplectic SP$\left( 2r,%
\mathbb{R}\right) $ family; in particular on the 5D $\mathcal{N}=1$ super
QFT with SP$\left( 4,\mathbb{R}\right) $ invariance. The first SP$\left( 2,%
\mathbb{R}\right) $ member is isomorphic to the 5D $\mathcal{N}=1$ SU$\left(
2\right) $ model of the unitary series. To achieve this goal, we $\left(
\mathbf{i}\right) $ revisit the toric Calabi-Yau threefold $\hat{X}\left(
Y^{4,0}\right) $ (p=4 and q=0), hosting a lifted SU$\left( 4\right) _{0}$
gauge symmetry; and $\left( \mathbf{ii}\right) $ reconsider the BPS quiver $%
Q_{\hat{X}\left( Y^{4,0}\right) }^{{\small SU}_{{\small 4}}}$ of the
underlying with 5D $\mathcal{N}=1$ super QFT compactified on a circle with
finite size. After that we develop an approach to construct toric Calabi-Yau
threefolds with symplectic symmetry and a method to build the BPS quiver $Q_{%
\hat{X}\left( Y^{4,0}\right) }^{{\small SP}_{{\small 4}}}$ with SP$\left( 4,%
\mathbb{R}\right) $ invariance. The extension of this construction to other
gauge symmetries is discussed in the conclusion section.\  \  \  \  \newline
The organisation is as follows: In section 2, we review properties of the
toric diagram $\Delta _{\hat{X}\left( Y^{4,0}\right) }^{{\small SU}_{{\small %
4}}}$ of the Calabi-Yau threefolds $\hat{X}\left( Y^{4,0}\right) $. We show
that $\Delta _{\hat{X}\left( Y^{4,0}\right) }^{{\small SU}_{{\small 4}}}$
has non trivial outer-automorphisms $H_{\Delta ^{{\small SU}_{{\small 4}%
}}}^{outer}$ having a fix point. We also show that this discrete group $%
H_{\Delta ^{{\small SU}_{{\small 4}}}}^{outer}$ can be interpreted as a
parity symmetry in $\mathbb{Z}^{2}$ lattice. In section 3, we investigate
the properties of the BPS quiver $Q_{\hat{X}\left( Y^{4,0}\right) }^{{\small %
SU}_{{\small 4}}}$ associated with $\Delta _{\hat{X}\left( Y^{4,0}\right) }^{%
{\small SU}_{{\small 4}}}$. Here we show that $Q_{\hat{X}\left(
Y^{4,0}\right) }^{{\small SU}_{{\small 4}}}$ has also an outer-automorphism
symmetry $H_{Q^{{\small SU}_{{\small 4}}}}^{outer}$ with fix points. This
outer-automorphism group has two factors given by $\left( \mathbb{Z}%
_{4}\right) _{Q^{{\small SU}_{{\small 4}}}}\times $ $\left( \mathbb{Z}%
_{2}^{outer}\right) _{Q^{{\small SU}_{{\small 4}}}}$. In section 4, we
introduce a new diagram to represent the toric $\hat{X}\left( Y^{4,0}\right)
.$ It is given by a graph $\mathcal{G}$ where the Calabi-Yau condition is
manifestly exhibited. To avoid confusion, we denote this graph like $%
\mathcal{G}_{\hat{X}\left( Y^{4,0}\right) }^{{\small SU}_{{\small 4}}}$ and
refer to it as the \emph{unitary CY graph} of the toric $\hat{X}\left(
Y^{4,0}\right) $ with SU$\left( 4\right) $ gauge symmetry. To deepen the
construction, we also give the unitary CY graphs $\mathcal{G}_{\hat{X}\left(
Y^{2,0}\right) }^{{\small SU}_{{\small 2}}}$ and $\mathcal{G}_{\hat{X}\left(
Y^{3,0}\right) }^{{\small SU}_{{\small 3}}}$ representing the toric $\hat{X}%
\left( Y^{p,0}\right) $ with p=2 and p=3. In section 5, we construct the
symplectic CY graph $\mathcal{\tilde{G}}_{\hat{X}\left( Y^{4,0}\right) }^{%
{\small SP}_{{\small 4}}}$ and the associated symplectic BPS quiver $\tilde{Q%
}_{\hat{X}\left( Y^{4,0}\right) }^{{\small SP}_{{\small 4}}}$. In section 6,
we give a conclusion and make comments. In the appendix, we give useful
properties on the geometric properties of the Coulomb branch of M-theory on
CY3s and describe the building of BPS quivers.

\section{Conical Sasaki-Einstein threefold $\hat{X}(Y^{4,0})$}

We begin by recalling that the Calabi-Yau threefold $\hat{X}(Y^{4,0}),$
taken as a real metric cone over the 5d Sasaki-Einstein variety \textrm{\cite%
{18h,19h}}, is a toric complex 3d manifold whose toric diagram $\Delta _{%
\hat{X}(Y^{4,0})}^{{\small SU}_{{\small 4}}}$ is a finite sublattice of $%
\mathbb{Z}^{3}$ as in the Figure \textbf{\ref{1}}. This toric $\Delta _{\hat{%
X}(Y^{4,0})}^{{\small SU}_{{\small 4}}}$ has seven points \textrm{given by}:
$\left( \mathbf{a}\right) $ Four external points $\mathbf{W}_{1},\mathbf{W}%
_{2},\mathbf{W}_{3},\mathbf{W}_{4}$ defining the \textrm{geometry on which
rests} the singularity of the SU$\left( 4\right) $ gauge fiber. $\left(
\mathbf{b}\right) $ Three internal points $\mathbf{V}_{1},\mathbf{V}_{2},%
\mathbf{V}_{3}$\ describing the crepant resolution of the singularity. This
resolution can be imagined as an intrinsic sub-geometry of the toric $\hat{X}%
(Y^{4,0})$ to which we often refer to as the fiber geometry.

\subsection{Toric diagram $\Delta _{\hat{X}(Y^{4,0})}^{{\protect \small SU}_{%
{\protect \small 4}}}$ and divisors of $\hat{X}(Y^{4,0})$}

\begin{figure}[tbph]
\begin{center}
\includegraphics[width=6cm]{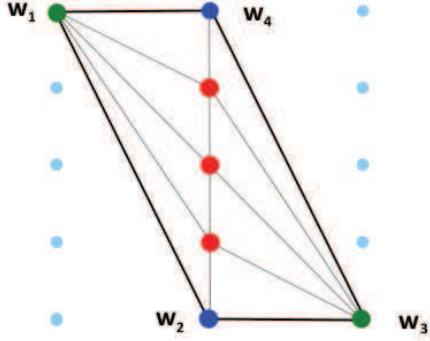}
\end{center}
\par
\vspace{-0.5cm}
\caption{The toric diagrams $\Delta _{\hat{X}(Y^{4,0})}^{{\protect \small SU}%
_{{\protect \small 4}}}$ having four external points (two blue and two green)
and three internal points (in red). These red points are associated with the
lifting of the SU$\left( 4\right) $ singularity of the gauge fiber. The
surface of $\Delta _{\hat{X}(Y^{4,0})}^{{\protect \small SU}_{{\protect \small %
4}}}$ is divided into 4+4 triangles. By merging the red points into the
first point, one is left with 4 triangles.}
\label{1}
\end{figure}

The four above mentioned $\mathbf{W}_{i}$ points (i=1,2,3,4) of the toric
diagram $\Delta _{\hat{X}(Y^{4,0})}^{{\small SU}_{{\small 4}}}$ can be also
interpreted as associated with four non compact divisors $D_{i}$ of the
toric $\hat{X}(Y^{4,0}).$ Similarly, the three internal points $\mathbf{V}%
_{a}$ ($a=1,2,3)$\ are interpreted as corresponding to three divisors $E_{a}$
of the toric $\hat{X}(Y^{4,0})$; but with the difference that the three $%
E_{a}$'s are compact complex 2d surfaces. In terms of the classes of these
divisors, the Calabi-Yau condition of the toric $\hat{X}(Y^{4,0})$ is given
by the vanishing sum; \textrm{see also appendix A},
\begin{equation}
\sum_{i=1}^{4}D_{i}+\sum_{a=1}^{3}E_{a}=0  \label{cyde}
\end{equation}%
This homological condition is implemented at the level of the toric diagram
by restricting the seven points of $\Delta _{\hat{X}(Y^{4,0})}^{{\small SU}_{%
{\small 4}}}$ to sit in the same hyperplane by taking the external like $%
\mathbf{W}_{i}=\left( \mathbf{w}_{i},1\right) $ and the internal points as $%
\mathbf{V}_{a}=\left( \mathbf{v}_{a},1\right) $ with $\mathbf{w}_{i}$ and $%
\mathbf{v}_{a}$ belonging to $\mathbb{Z}^{2}$. A particular realisation of
the seven points of $\Delta _{\hat{X}(Y^{4,0})}^{{\small SU}_{{\small 4}}}$
is given by
\begin{table}[h]
\centering \renewcommand{\arraystretch}{1.2} $%
\begin{tabular}{|l|l|l|l|l||l|ll|}
\hline
$\  \Delta _{\hat{X}(Y^{4,0})}$ & $\  \  \ w_{1}$ & $\  \ w_{2}$ & $\  \ w_{3}$ &
$\  \ w_{4}$ & $\  \  \upsilon _{1}$ & $\  \  \upsilon _{2}$ & $\  \  \upsilon _{3}$
\\ \hline \hline
\ points & $\left( -1,4\right) $ & $\left( 0,0\right) $ & $\left( 1,0\right)
$ & $\left( 0,4\right) $ & $\left( 0,1\right) $ & $\left( 0,2\right) $ & $%
\left( 0,3\right) $ \\ \hline
divisors & $\  \  \  \  \ D_{1}$ & $\  \ D_{2}$ & $\  \ D_{3}$ & $\  \ D_{4}$ & $\
\ E_{1}$ & $\  \ E_{2}$ & $\  \ E_{3}$ \\ \hline \hline
\end{tabular}%
$%
\caption{Toric data of the\ Calabi-Yau threefold $\hat{X}\left(
Y^{4,0}\right) .$}
\label{pq}
\end{table}
The toric diagram $\Delta _{\hat{X}(Y^{4,0})}$ representing the resolved
Calabi-Yau threefold $\hat{X}(Y^{4,0})$ is depicted by the Figure \textbf{%
\ref{1} }where a triangulation the surface of $\Delta _{\hat{X}(Y^{4,0})}$
is highlighted \textrm{\cite{ds}}. It describes the lifting of the A$%
_{3}\simeq SU\left( 4\right) $ singularity\textbf{.}
Notice that Table \textrm{\textbf{\ref{pq}}} is the data for $\left(
p,q\right) =\left( 4,0\right) $ saturating the lower bound of the constraint
$0\leq q\leq 4.$ For\ generic values of q constrained like $0\leq q\leq p$,
we have the following data
\begin{table}[h]
\centering \renewcommand{\arraystretch}{1.2} $%
\begin{tabular}{|l|l|l|l|l||l|}
\hline
$\  \Delta _{\hat{X}(Y^{p,q})}$ & $\  \  \ w_{1}$ & $\  \ w_{2}$ & $\  \ w_{3}$ &
$\  \ w_{4}$ & $\  \  \left \{ \upsilon _{a}\right \} _{1\leq a\leq p-1}$ \\
\hline \hline
\ points & $\left( -1,p-q\right) $ & $\left( 0,0\right) $ & $\left(
1,0\right) $ & $\left( 0,p\right) $ & $\  \  \  \left( 0,a\right) $ \\ \hline
divisors & $\  \  \  \  \ D_{1}$ & $\  \ D_{2}$ & $\  \ D_{3}$ & $\  \ D_{4}$ & $\
\  \left \{ E_{a}\right \} _{1\leq a\leq p-1}$ \\ \hline \hline
\end{tabular}%
$%
\caption{Toric data of $\hat{X}\left( Y^{p,q}\right) $ with $0\leq q\leq p.$}
\label{m}
\end{table}
This toric $\Delta _{\hat{X}(Y^{p,q})}$ has 3+p points and then 3+p
divisors; p-1 of them are compact. They concern the divisor set $\left \{
E_{a}\right \} _{1\leq a\leq p-1}$. Notice also that the three internal
(red) points of $\Delta _{\hat{X}(Y^{4,0})}$\ represented by the Figure
\textbf{\ref{1}} form a (vertical) linear chain A$_{3}$ in the toric diagram
with boundary points effectively given by the two (blue) external $\mathbf{w}%
_{2}$ and $\mathbf{w}_{4}$. For convenience, we rename these two particular
boundary points like $\mathbf{w}_{2}=\mathbf{\upsilon }_{0}$ and $\mathbf{w}%
_{4}=\mathbf{\upsilon }_{4}$ so that the above mentioned chain A$_{3}$ can
be put in correspondence with the standard A$_{3}$- geometry of the ALE
space with resolved SU$\left( 4\right) $ singularity \textrm{\cite{5f,50f}}.
With this renaming, the Table \textbf{\ref{pq}} gets mapped to
\begin{table}[h]
\centering \renewcommand{\arraystretch}{1.2} $%
\begin{tabular}{|l||l|l|l|l|l||l|l||}
\hline
$\hat{X}(Y^{4,0})$ & \multicolumn{5}{||l}{\  \  \  \  \  \  \  \  \  \  \  \  \  \  \  \  \ A%
$_{3}$- geometry} & \multicolumn{2}{||l||}{transverse geometry} \\ \hline
$\  \Delta _{\hat{X}(Y^{4,0})}^{SU_{4}}$ & $\  \  \upsilon _{0}$ & $\  \
\upsilon _{1}$ & $\  \  \upsilon _{2}$ & $\  \  \upsilon _{3}$ & $\  \  \upsilon
_{4}$ & $\  \  \ w_{1}$ & $\  \ w_{3}$ \\ \hline \hline
\ points & $\left( 0,0\right) $ & $\left( 0,1\right) $ & $\left( 0,2\right) $
& $\left( 0,3\right) $ & $\left( 0,4\right) $ & $\left( -1,4\right) $ & $%
\left( 1,0\right) $ \\ \hline \hline
\end{tabular}%
$%
\caption{{}Toric data of the A$_{3}$- geometry within the resolved
Calabi-Yau threefold $\hat{X}(Y^{4,0}).$}
\label{q}
\end{table}
A similar description can be done for $\Delta _{\hat{X}(Y^{p,0})}.$ For
simplicity of the presentation, we omit it. Having introduced the particular
toric diagram $\Delta _{\hat{X}(Y^{4,0})}^{{\small SU}_{{\small 4}}}$
hosting an underlying unitary SU$\left( 4\right) $ gauge symmetry, we turn
now to explore one of its exotic properties namely its outer-automorphism
symmetries.

\subsection{Outer- automorphisms of $\Delta _{\hat{X}(Y^{4,0})}^{%
{\protect \small SU}_{{\protect \small 4}}}$}

\begin{figure}[tbph]
\begin{center}
\includegraphics[width=14cm]{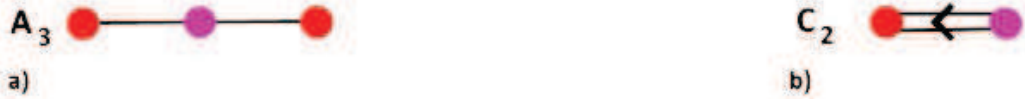}
\end{center}
\par
\vspace{-0.5cm}
\caption{\textbf{a)} The Dynkin diagram of the Lie algebra A$_{3}.$ It has a
mirror $\left( \mathbb{Z}_{2}\right) _{\Delta ^{SU_{4}}}$ outer-automorphism
symmetry leaving one node fixed (in magenta color). \textbf{b)} The Dynkin
diagram of the symplectic Lie algebra C$_{2}$. It is obtained by folding A$%
_{3}$ under $\left( \mathbb{Z}_{2}\right) _{\Delta ^{SU_{4}}}$.}
\label{A3}
\end{figure}

\QTP{Dialog Text}
A careful inspection of the Figure \textbf{\ref{1}} reveals that the toric
diagram $\Delta _{\hat{X}(Y^{4,0})}^{{\small SU}_{{\small 4}}}$ has
outer-automorphism symmetries forming a discrete group $H_{\Delta ^{{\small %
SU}_{4}}}^{outer}$. This is a finite symmetry group generated by the
following transformations of the external points $\mathbf{w}_{i}$ and the
internal $\mathbf{\upsilon }_{a}$s,
\begin{equation}
H_{\Delta ^{{\small SU}_{4}}}^{outer}:\mathbf{w}_{1}\leftrightarrow \mathbf{w%
}_{3},\qquad \mathbf{w}_{2}\leftrightarrow \mathbf{w}_{4},\qquad \mathbf{%
\upsilon }_{a}\leftrightarrow \mathbf{\upsilon }_{4-a}
\end{equation}%
Notice that the outer-automorphisms in the gauge fiber act by exchanging the
two internal $\mathbf{\upsilon }_{1}\leftrightarrow \mathbf{\upsilon }_{3}$;
but fix the central point $\mathbf{\upsilon }_{2}.$ This property is
interesting; it will be used later on to engineer a new \textrm{gauge }%
fiber. By using the \textrm{parametrisation} $\mathbf{w}_{i}=\left(
w_{i}^{x},w_{i}^{y}\right) $ and $\mathbf{v}_{a}=\left(
v_{a}^{x},v_{a}^{y}\right) $, we learn that the outer-automorphism group $%
H_{\Delta ^{{\small SU}_{4}}}^{outer}$ is given by the product of two
reflections like%
\begin{equation}
H_{\Delta ^{{\small SU}_{4}}}^{outer}=\left( \mathbb{Z}_{2}^{x}\right)
_{\Delta ^{{\small SU}_{4}}}\times \left( \mathbb{Z}_{2}^{y}\right) _{\Delta
^{{\small SU}_{4}}}  \label{su4}
\end{equation}%
with
\begin{equation}
\begin{tabular}{lllllll}
$\left( \mathbb{Z}_{2}^{x}\right) _{\Delta ^{{\small SU}_{4}}}$ & $:$ & $%
w_{1}^{x}\rightarrow -w_{1}^{x}$ & , & $w_{3}^{x}\rightarrow -w_{3}^{x}$ & ,
& $v_{a}^{x}\rightarrow -v_{a}^{x}$ \\
$\left( \mathbb{Z}_{2}^{y}\right) _{\Delta ^{{\small SU}_{4}}}$ & $:$ & $%
w_{1}^{y}\rightarrow w_{3}^{y}$ & , & $w_{3}^{y}\rightarrow w_{1}^{y}$ & , &
$v_{a}^{y}\rightarrow v_{4-a}^{y}$%
\end{tabular}%
\end{equation}%
Form these outer-automorphism transformations, we learn that $\left( \mathbb{%
Z}_{2}^{x}\right) _{\Delta ^{{\small SU}_{4}}}$ acts trivially on the
internal points $\mathbf{v}_{a}$ of the A$_{3}$- linear chain of $\Delta _{%
\hat{X}(Y^{4,0})}^{{\small SU}_{{\small 4}}}.$ So the group $\left( \mathbb{Z%
}_{2}^{x}\right) _{\Delta ^{{\small SU}_{4}}}$ leaves invariant the A$_{3}$-
gauge fiber within the toric Calabi-Yau $\hat{X}(Y^{4,0}).$ It affects only
the external points $\mathbf{w}_{1}$ and $\mathbf{w}_{3}$ which are
associated with the transverse geometry shown in the table \textbf{\ref{q}}.
Regarding the $\left( \mathbb{Z}_{2}^{y}\right) _{\Delta ^{{\small SU}_{4}}}$
reflection, it acts non trivially on the points of the A$_{3}$-chain; we
have:
\begin{equation}
\left( \mathbb{Z}_{2}^{y}\right) _{\Delta ^{{\small SU}_{4}}}:\mathbf{%
\upsilon }_{a}\rightarrow \mathbf{\upsilon }_{4-a}
\end{equation}%
Under this mirror symmetry, the A$_{3}$- gauge fiber has then a fix point
which is an interesting feature that we want to exploit to build a new gauge
fiber by using folding ideas \textrm{\cite{M,0M,13h,14h}}. In this regards,
recall that the $\left( \mathbb{Z}_{2}^{y}\right) _{\Delta ^{{\small SU}%
_{4}}}$ action looks like a well known outer-automorphism symmetry group $%
\mathbb{Z}_{2}$ that we encounter in the folding of the Dynkin diagrams of
the finite dimensional Lie algebras $A_{2r-1}$. Here, we are dealing with
the particular $A_{3}\sim SU\left( 4\right) $ which is just the leading non
trivial member of the $A_{2r-1}$ series. As an illustration; see the
pictures of the Figure \textbf{\ref{A3}} describing the folding of the
Dynkin diagram $A_{3}$ giving the Dynkin diagram of the symplectic C$%
_{2}\simeq sp\left( 4,\mathbb{R}\right) $ which, thought not relevant for
our present study, it is also isomorphic to B$_{2}\simeq so\left( 5\right) $%
.
Recall as well that the Dynkin diagrams of finite dimensional Lie algebras $%
\mathbf{g}$ may be also thought of in terms of the Cartan matrices $K\left(
\mathbf{g}\right) _{ij}=\mathbf{\alpha }_{i}^{\vee }.\mathbf{\alpha }_{j}$
defined by the intersection of simple roots $\mathbf{\alpha }_{i}$ and
co-roots $\mathbf{\alpha }_{i}^{\vee }=\mathbf{\alpha }_{i}/\mathbf{\alpha }%
_{i}^{2}$. For the examples of $A_{3}\simeq su\left( 4\right) $ and $%
C_{2}\simeq sp\left( 4,\mathbb{R}\right) ,$ we have the following matrices%
\begin{equation}
K\left( A_{3}\right) =\left(
\begin{array}{ccc}
2 & -1 & 0 \\
-1 & 2 & -1 \\
0 & -1 & 2%
\end{array}%
\right) \qquad ,\qquad K\left( C_{2}\right) =\left(
\begin{array}{cc}
2 & -1 \\
-2 & 2%
\end{array}%
\right)
\end{equation}%
Notice that the picture on the left of the Figure \textbf{\ref{A3}} can be
put in correspondence with the internal (red) points of the A$_{3}$- linear
chain of the Figure \textbf{\ref{1}. }At this level, one may ask what about
toric diagrams with a C$_{2}$ type sub-diagram. We will answer this question
later on after highlighting another property of $\Delta _{\hat{X}(Y^{4,0})}^{%
{\small SU}_{{\small 4}}}$. Before that, let us describe succinctly the BPS
quivers associated with the toric diagram of $\hat{X}(Y^{4,0})$; and study
its outer-automorphisms.

\section{BPS quiver $Q_{\hat{X}(Y^{p,0})}^{{\protect \small SU}_{%
{\protect \small p}}}:$ cases $p=3,4$}

In this section, we investigate two examples of unitary BPS quivers namely
the $Q_{\hat{X}(Y^{3,0})}^{{\small SU}_{{\small 3}}}$ and the $Q_{\hat{X}%
(Y^{4,0})}^{{\small SU}4}.$ These unitary BPS quivers are representatives of
the families $Q_{\hat{X}(Y^{2r-1,0})}^{{\small SU}_{{\small 2r-1}}}$ and $Q_{%
\hat{X}(Y^{2r,0})}^{{\small SU}_{{\small 2r}}}$ with $r\geq 1.$ They have
intrinsic properties that we want to study and which will be used later on.
First, we consider the quiver $Q_{\hat{X}(Y^{4,0})}^{{\small SU}_{{\small 4}%
}}$ with gauge symmetry SU$\left( 4\right) $ as this quiver is one of the
main graphs that interests us in this study. Then, we turn to the BPS quiver
$Q_{\hat{X}(Y^{3,0})}^{{\small SU}_{{\small 3}}}$ with unitary symmetry SU$%
\left( 3\right) $. The $Q_{\hat{X}(Y^{3,0})}^{{\small SU}_{{\small 3}}}$
quiver is reported here for a matter of comparison with $Q_{\hat{X}%
(Y^{4,0})}^{{\small SU}_{{\small 4}}}.$ The results obtained for these
quivers hold as well for the families $Q_{\hat{X}(Y^{2r-1,0})}^{{\small SU}_{%
{\small 2r-1}}}$ and $Q_{\hat{X}(Y^{2r,0})}^{{\small SU}_{{\small 2r}}}$.

\subsection{BPS quiver $Q_{\hat{X}(Y^{4,0})}^{{\protect \small SU}_{%
{\protect \small 4}}}$}

\begin{figure}[tbph]
\begin{center}
\includegraphics[width=6cm]{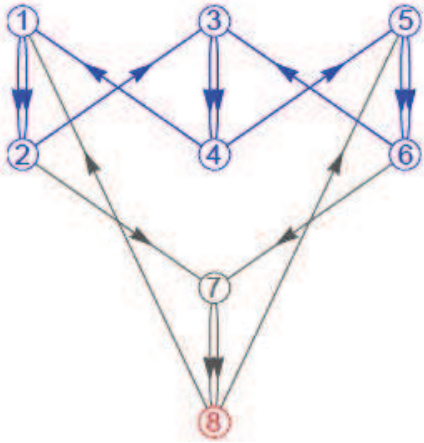}
\end{center}
\par
\vspace{-0.5cm}
\caption{BPS quiver $Q_{\hat{X}(Y^{4,0})}^{{\protect \small SU}_{%
{\protect \small 4}}}$ associated with $\hat{X}\left( Y^{4,0}\right) .$ Its
elementary BPS states are represented by the 8 nodes and are linked by 16
edges. The shape of this BPS quiver has been borrowed from the paper
\protect \cite{3e} ---Figure 25-(a)---. The subdiagram with 6 blue nodes and\
10 blue arrows refer to the 4D subquiver.}
\label{2}
\end{figure}

The construction of the unitary BPS quiver $Q_{\hat{X}(Y^{4,0})}^{{\small SU}%
_{{\small 4}}}$ of the 5D $\mathcal{N}=1$ super QFTs, compactified on
\textrm{a circle} with finite size and based on $\hat{X}(Y^{4,0})$, follows
from the brane tiling of the so called brane-web $\tilde{\Delta}_{\hat{X}%
(Y^{4,0})}^{{\small SU}_{{\small 4}}}$ (the dual\ of the toric diagram $%
\Delta _{\hat{X}(Y^{4,0})}^{{\small SU}_{{\small 4}}}$) by applying the fast
inverse algorithm \textrm{\cite{1g,2g,3g}}. Up to a Seiberg- type duality
transformation, the representative $Q_{\hat{X}(Y^{4,0})}^{{\small SU}_{%
{\small 4}}}$ has a quiver- dimension $\boldsymbol{d}_{bps}$ equals to $%
2\times 3+2.$ Then the $Q_{\hat{X}(Y^{4,0})}^{{\small SU}_{{\small 4}}}$ has
8 elementary BPS particles that generate the BPS spectrum of the 5D super
QFT. \textrm{For further details; see the appendices A and B}. To fix ideas,
\textrm{let us illustrate the numbers} involved in the $\boldsymbol{d}_{bps}$
\textrm{dimension} which for $Q_{\hat{X}(Y^{p,0})}^{{\small SU}_{{\small p}%
}} $ with p$\geq $2 reads as follows; \textrm{see also eqs(\ref{nf}-\ref{fn})%
},
\begin{equation}
\boldsymbol{d}_{bps}=2\left( p-1\right) +2  \label{dps}
\end{equation}%
$\left( \mathbf{i}\right) $ the number $3=4-1$ is precisely the rank of the
SU$\left( 4\right) $ gauge fiber within the toric $\hat{X}(Y^{4,0}).$ It is
also the number of compact divisors ---$E_{1},E_{2},E_{3}$--- of the
threefold $\hat{X}(Y^{4,0})$. $\left( \mathbf{ii}\right) $ The product $%
2\times 3=6$ designates the number of the electric/magnetic charged
particles. These 3+3 particles have interpretation in terms of M2-- and
M5-branes wrapping 2- and 4- cycles in the internal threefold $\hat{X}%
(Y^{4,0})$. $\left( \mathbf{iii}\right) $ The extra number 2=1+1 in the $%
\boldsymbol{d}_{bps}$- dimension of $Q_{\hat{X}(Y^{4,0})}^{{\small SU}_{%
{\small 4}}}$ refers to an instanton and to the elementary Kaluza-Klein D0
brane; for more details see \textrm{\cite{1e,2e,3e} and the appendix A}.
\newline
The schematic structure of the BPS quiver $Q_{\hat{X}(Y^{4,0})}^{{\small SU}%
_{{\small 4}}}$ is depicted by the Figure \textbf{\ref{2}}.
It has 8 nodes $\left \{ 1\right \} ,...,\left \{ 8\right \} $ interpreted
in terms of 8 elementary BPS particles. These nodes organise into four
Kronecker quivers (4 doublets of nodes) denoted like $\kappa _{c}=\left \{
2c-1,2c\right \} _{1\leq c\leq 4}$; explicitly, we have:%
\begin{equation}
\begin{tabular}{lllllll}
$\kappa _{1}$ & $=$ & $\left \{ 1,2\right \} $ & , & $\kappa _{3}$ & $=$ & $%
\left \{ 5,6\right \} $ \\
$\kappa _{2}$ & $=$ & $\left \{ 3,4\right \} $ & , & $\kappa _{4}$ & $=$ & $%
\left \{ 7,8\right \} $%
\end{tabular}%
\end{equation}%
As shown by the Figure \textbf{\ref{2}}, the 8 nodes of the BPS quiver are
linked by $4\times 4=16$ quiver- edges $\left \langle j|l\right \rangle $
interpreted in terms of chiral superfields in the language of supersymmetric
quantum mechanics (SQM) \textrm{\cite{2e}}. The unitary BPS quiver $Q_{\hat{X%
}(Y^{4,0})}^{{\small SU}_{{\small 4}}}$ \textrm{has} been first considered
in \textrm{\cite{3e} (see figure 25-a, page 61). F}or later use, we re-draw
the Figure \textbf{\ref{2}} as depicted by the equivalent Figure \textbf{\ref%
{3}.}
\begin{figure}[tbph]
\begin{center}
\includegraphics[width=14cm]{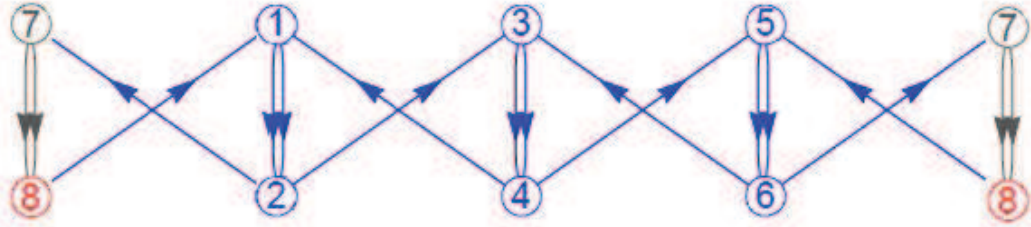}
\end{center}
\par
\vspace{-0.5cm}
\caption{The BPS quiver is $Q_{\hat{X}(Y^{4,0})}^{{\protect \small SU}_{%
{\protect \small 4}}}$ has eight nodes organised into 4 vertical pairs $%
\left
\{ 2a-1;2a\right \} .$ In this representative chain, the first pair
and the last one are given by $\left \{ 7;8\right \} .$ They should be
identified as they concern the same nodes' pair.}
\label{3}
\end{figure}
In this redrawing\textbf{, }we have represented the Kronecker quiver $%
\left
\{ 7,8\right \} $ twice. This way of doing allows to think of the $Q_{%
\hat{X}(Y^{4,0})}^{{\small SU}_{{\small 4}}}$ as a periodic chain of
Kronecker quivers with periodicity generated by a $\left( \mathbb{Z}%
_{4}\right) _{Q^{{\small SU}_{{\small 4}}}}$ outer-automorphism symmetry
acting $\left( \mathbf{i}\right) $ on the quiver- nodes $\left \{
2c-1\right
\} _{1\leq c\leq 4}$ and $\left \{ 2c\right \} _{1\leq c\leq 4}$
as follows%
\begin{equation}
\left( \mathbb{Z}_{4}\right) _{Q^{{\small SU}_{{\small 4}}}}:\left.
\begin{array}{ccc}
\left \{ 2c-1\right \} & \rightarrow & \left \{ 2c+7\right \} \\
\text{ \  \  \  \ }\left \{ 2c\right \} & \rightarrow & \left \{ 2c+8\right \}%
\end{array}%
\right.  \label{z4}
\end{equation}%
and $\left( \mathbf{ii}\right) $ on the Kronecker quivers like $\kappa
_{c}\rightarrow \kappa _{c+4}$. These outer-automorphisms, which act also on
the oriented arrows, have no fix node and no fix arrow. They play a
secondary role in our construction.\newline
In addition to $\left( \mathbb{Z}_{4}\right) _{Q^{{\small SU}_{{\small 4}}}}$%
, the unitary BPS quiver $Q_{\hat{X}(Y^{4,0})}^{{\small SU}_{{\small 4}}}$
has another outer-automorphism group factor namely $\left( \mathbb{Z}_{2}^{%
{\small outer}}\right) _{Q^{{\small SU}_{{\small 4}}}}$. It acts as a
reflection symmetry mirroring nodes and exchanging oriented arrows. Contrary
to $\left( \mathbb{Z}_{4}\right) _{Q^{{\small SU}_{{\small 4}}}}$, the
mirror $\left( \mathbb{Z}_{2}^{{\small outer}}\right) _{Q^{{\small SU}_{%
{\small 4}}}}$ has the remarkable property of fixing four quiver- nodes and
the associated arrows. It acts on the Kronecker quivers as%
\begin{equation}
\left( \mathbb{Z}_{2}^{{\small outer}}\right) _{Q^{{\small SU}_{{\small 4}%
}}}:\kappa _{c}\rightarrow \kappa _{4-c}  \label{z2}
\end{equation}%
thus exchanging $\kappa _{1}\leftrightarrow \kappa _{3}$; but fixing $\kappa
_{2}$ and $\kappa _{4}$ since $\kappa _{4}\equiv \kappa _{0}$ due to the
periodicity property $\kappa _{c}\simeq \kappa _{c+4};$ thanks to $\left(
\mathbb{Z}_{4}\right) _{Q^{{\small SU}_{{\small 4}}}}$. By denoting the
eight nodes like $\left \{ 1\right \} ,...,\left \{ 8\right \} ,$ the $%
\left( \mathbb{Z}_{2}^{{\small outer}}\right) _{Q^{{\small SU}_{{\small 4}%
}}} $ group is then generated by the double transposition $\left( \left \{
1\right \} \left \{ 5\right \} \right) \circ \left( \left \{ 2\right \}
\left \{ 6\right \} \right) $. Below, we refer to this double transposition
simply as $\left( 15\right) \left( 26\right) $; so, we have:%
\begin{equation}
\left( \mathbb{Z}_{2}^{{\small outer}}\right) _{Q^{{\small SU}_{{\small 4}%
}}}=\left \{ s=\left( 15\right) \left( 26\right) \quad |\quad s^{2}=I\right
\}
\end{equation}%
From this description, \textrm{we learn two interesting} things: First, the $%
\left( \mathbb{Z}_{2}^{{\small outer}}\right) _{Q^{{\small SU}_{{\small 4}%
}}} $ is a particular subgroup of the symmetric (permutation) group $\mathbb{%
S}_{8}$ of eight elements (nodes) $\left \{ 1,...,8\right \} .$ Second the $%
\left( \mathbb{Z}_{4}\right) _{Q^{{\small SU}_{{\small 4}}}}$ is also a
subgroup of $\mathbb{S}_{8}$; it generated by the product of two 4-cycles as
follows,%
\begin{equation}
\left( \mathbb{Z}_{4}\right) _{Q^{{\small SU}_{{\small 4}}}}=\left \{
t=\left( 1357\right) \left( 2468\right) \quad |\quad t^{4}=I\right \}
\end{equation}%
So, both $\left( \mathbb{Z}_{2}^{{\small outer}}\right) _{Q^{{\small SU}_{%
{\small 4}}}}$ and $\left( \mathbb{Z}_{4}\right) _{Q^{{\small SU}_{{\small 4}%
}}}$ are subgroups of the enveloping $\mathbb{S}_{8}$. Similar
outer-automorphism groups can be written down for the family $Q_{\hat{X}%
(Y^{2r,0})}^{{\small SU}_{{\small 2r}}}$ with $r\geq 2$.

\subsection{BPS quiver $Q_{\hat{X}(Y^{3,0})}^{{\protect \small SU}_{%
{\protect \small 3}}}$}

\QTP{Dialog Text}
Here, we study the BPS quiver $Q_{\hat{X}(Y^{3,0})}^{{\small SU}_{{\small 3}%
}}$ and some of its outer-automorphisms in order to compare with $Q_{\hat{X}%
(Y^{4,0})}^{{\small SU}_{{\small 4}}}.$ The BPS quiver $Q_{\hat{X}%
(Y^{3,0})}^{{\small SU}_{{\small 3}}}$ with gauge symmetry SU$\left(
3\right) $ has a quite similar structure as $Q_{\hat{X}(Y^{4,0})}^{{\small SU%
}_{{\small 4}}}$; but a different quiver dimension which is given by
\begin{equation}
\boldsymbol{d}_{bps}=2\left( p-1\right) +2=2\times 2+2=6
\end{equation}%
As such, the BPS quiver $Q_{\hat{X}(Y^{3,0})}^{{\small SU}_{{\small 3}}}$
has six nodes $\left \{ 1\right \} ,...,\left \{ 6\right \} $ interpreted in
terms of 6 elementary BPS particles. They organise into three Kronecker
quivers \textrm{namely}
\begin{equation}
\begin{tabular}{lllllllllll}
$\kappa _{1}$ & $=$ & $\left \{ 1,2\right \} $ & , & $\kappa _{2}$ & $=$ & $%
\left \{ 3,4\right \} $ & , & $\kappa _{3}$ & $=$ & $\left \{ 5,6\right \} $%
\end{tabular}%
\end{equation}%
This BPS quiver has 12 oriented arrows as depicted by the Figure \textbf{\ref%
{q30}}.
\begin{figure}[tbph]
\begin{center}
\includegraphics[width=4cm]{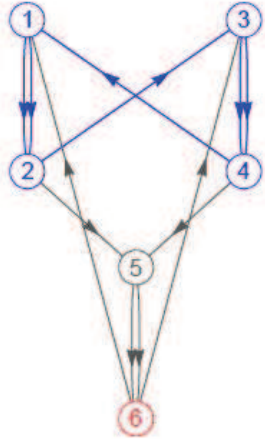}
\end{center}
\par
\vspace{-0.5cm}
\caption{BPS quiver $Q_{\hat{X}(Y^{3,0})}^{{\protect \small SU}_{%
{\protect \small 3}}}$ associated with $\hat{X}\left( Y^{3,0}\right) .$ Its
elementary BPS states are represented by the 6 nodes and are linked by 12
edges. The shape of this BPS quiver has been borrowed from the paper
\protect \cite{3e} ---Figure 22-(a) page 58---. The subgraph with four blue
points and 6 orientied arrows correspond to the quiver of the 4D super QFT.}
\label{q30}
\end{figure}
Though not very important for our present study as it cannot induce a BPS
quiver with symplectic gauge symmetry, notice that the quiver $Q_{\hat{X}%
(Y^{3,0})}^{{\small SU}_{{\small 3}}}$ has also outer-automorphism
symmetries forming a group $H_{Q^{{\small SU}_{{\small 3}}}}^{{\small outer}%
} $ with two factors as given below
\begin{equation}
H_{Q^{{\small SU}_{{\small 3}}}}^{{\small outer}}=\left( \mathbb{Z}_{2}^{%
{\small outer}}\right) _{Q^{{\small SU}_{{\small 3}}}}\times \left( \mathbb{Z%
}_{3}\right) _{Q^{{\small SU}_{{\small 3}}}}
\end{equation}%
The factor $\left( \mathbb{Z}_{2}^{{\small outer}}\right) _{Q^{{\small SU}_{%
{\small 3}}}}$ exchanges the nodes $\left \{ 1\right \} \leftrightarrow
\left \{ 3\right \} $ and $\left \{ 2\right \} \leftrightarrow \left \{
4\right \} $; but it fixes the nodes $\left \{ 5\right \} $ and $\left \{
6\right \} $ as clearly seen on the Figure \textbf{\ref{q30}}. In terms of
the Kronecker quivers, we have $\kappa _{1}\leftrightarrow \kappa _{2}$ but $%
\kappa _{3}\leftrightarrow \kappa _{3}$. So, the outer-automorphisms of $Q_{%
\hat{X}(Y^{3,0})}^{{\small SU}_{{\small 3}}}$ are different from those of
the quiver $Q_{\hat{X}(Y^{4,0})}^{{\small SU}_{{\small 4}}}$ which are given
by Figure \textbf{\ref{3}.} Recall that\textbf{\ }the $H_{Q^{{\small SU}_{%
{\small 4}}}}^{{\small outer}}$ has\textbf{\ }four fix nodes instead of two
for $Q_{\hat{X}(Y^{3,0})}^{{\small SU}_{{\small 3}}}$; i.e: two Kronecker
quivers for $H_{Q^{{\small SU}_{{\small 4}}}}^{{\small outer}}$, against one
Kronecker quiver for $H_{Q^{{\small SU}_{{\small 3}}}}^{{\small outer}}$%
\textbf{. }This difference holds as well for generic quivers $Q_{\hat{X}%
(Y^{2r,0})}^{{\small SU}_{{\small 2r}}}$ and $Q_{\hat{X}(Y^{2r-1,0})}^{%
{\small SU}_{{\small 2r-1}}}$ with respective outer-automorphism groups $%
H_{Q^{{\small SU}_{{\small 2r}}}}^{{\small outer}}$ and $H_{Q^{{\small SU}_{%
{\small 2r-1}}}}^{{\small outer}}.$\newline
Regarding the\textbf{\ }factor $\left( \mathbb{Z}_{3}\right) _{Q^{{\small SU}%
_{{\small 3}}}}$, it allows to represent the quiver $Q_{\hat{X}(Y^{3,0})}^{%
{\small SU}_{{\small 3}}}$ as a periodic chain as depicted by the Figure
\textbf{\ref{q300}.}
\begin{figure}[tbph]
\begin{center}
\includegraphics[width=8cm]{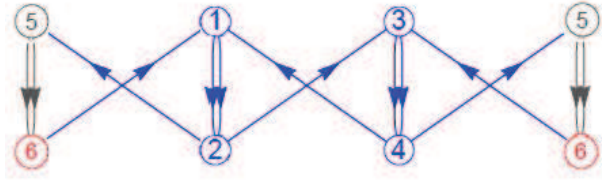}
\end{center}
\par
\vspace{-0.5cm}
\caption{BPS quiver $Q_{\hat{X}(Y^{3,0})}^{{\protect \small SU}_{%
{\protect \small 3}}}$ associated with $\hat{X}\left( Y^{3,0}\right) .$ Its
elementary BPS states are represented by the 6 nodes and are linked by 12
edges. The outer-automorphism group $\left( \mathbb{Z}_{2}^{{\protect \small %
outer}}\right) _{Q^{{\protect \small SU}_{{\protect \small 3}}}}$ has no fix
point.}
\label{q300}
\end{figure}

\section{Graphs $\mathcal{G}_{\hat{X}(Y^{p,q})}$ with manifest CY condition}

In this section, we \textrm{introduce a new graph} to deal with the toric
diagram $\Delta _{\hat{X}(Y^{4,0})}^{{\small SU}_{{\small 4}}}$ with p=4
representing the Calabi-Yau threefold $\hat{X}(Y^{4,0})$ with a resolved SU$%
\left( 4\right) $ gauge fiber. We refer to this \textrm{new graph} as \emph{%
the} \emph{unitary Calabi-Yau graph} and we denote it like $\mathcal{G}_{%
\hat{X}(Y^{4,0})}^{{\small SU}_{{\small 4}}}.$ \textrm{This graph} is
explicitly defined by $p-1$ vector $\mathbf{q}^{b}$ with components given by
the triple intersection numbers%
\begin{equation}
q_{A}^{b}=\mathcal{D}_{A}.E_{b}^{2}  \label{fr}
\end{equation}%
where the label $A=\left( i,a\right) $ with $i=1,2,3,4,$ for non compact
divisors D$_{i},$ and $a=1,2,3$ for the compact $E_{a}$. Below, we refer to
these $\mathbf{q}^{b}$'s as generalised Mori-vectors. Though this CY graph $%
\mathcal{G}_{\hat{X}(Y^{4,0})}^{{\small SU}_{{\small 4}}}$ looks formally
different from the toric diagram, it is in fact equivalent to it. It is just
another way to deal with $\Delta _{\hat{X}(Y^{4,0})}^{{\small SU}_{{\small 4}%
}}$ where the Calabi-Yau condition is manifestly exhibited. As we will show
below, this is useful in looking for solutions of underlying constraint
relations required by the toric threefold $\hat{X}(Y^{4,0})$.

\subsection{Building the CY graph $\mathcal{G}_{\hat{X}(Y^{4,0})}^{%
{\protect \small SU}_{{\protect \small 4}}}$}

\begin{figure}[tbph]
\begin{center}
\includegraphics[width=6cm]{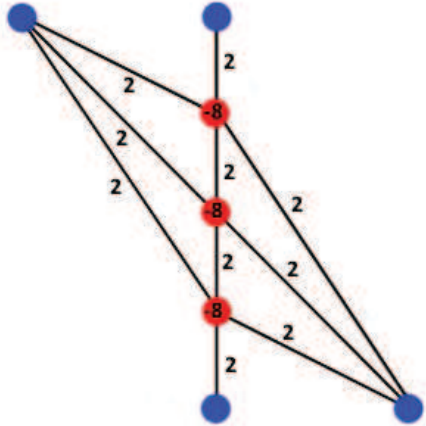}
\end{center}
\par
\vspace{-0.5cm}
\caption{The graph of the CY threefolds geometry $\hat{X}\left(
Y^{4,0}\right) $ exhibiting manifestly the CY condition at each internal
point. This graph has three (red) compact 4-cycles E$_{1}$,E$_{2}$,E$_{3}$,
each with triple self intersection $\left( -8\right) $, intersecting four
non compact (blue) 4-cycles. The Calabi-Yau threefolds condition is ensured
by the vanishing sum of the total charge at each red exceptional node. The
underlying SQFT has an SU$\left( 4\right) $ gauge symmetry. Notice also that
this graph has a remarkable outer-automorphism symmetry to used later on.}
\label{y4-geo}
\end{figure}

To engineer the unitary Calabi-Yau graph $\mathcal{G}_{\hat{X}(Y^{4,0})}^{%
{\small SU}_{{\small 4}}}$ of the toric $\hat{X}(Y^{4,0}),$ we start form
the Calabi-Yau condition given by eq(\ref{cyde}) namely $%
\sum_{i=1}^{4}D_{i}+\sum_{a=1}^{3}E_{a}=0$. This constraint relation is
expressed in terms of the four non compact divisors D$_{i}$ and the three
compact E$_{a}$; but it is not the only constraint that must be obeyed by
the divisors. There \textrm{are two other constraints that must be satisfied
by the divisors}. So, the seven divisors $\left( D_{i},E_{a}\right) $ of the
toric Calabi-Yau threefolds $\hat{X}(Y^{4,0})$ are subject to three basic
constraints. They can be collectively expressed as 3-vector equation like
\begin{equation}
\sum_{i=1}^{4}\mathbf{W}_{i}D_{i}+\sum_{a=1}^{3}\mathbf{V}_{a}E_{a}=0
\end{equation}%
where $\mathbf{W}_{i}=\left( \mathbf{w}_{i},1\right) $ and $\mathbf{V}%
_{a}=\left( \mathbf{v}_{a},1\right) $ are as in Table \textbf{\ref{pq}}. To
deal with the CY constraint eq(\ref{cyde}), we bring it to a relation
between triple intersection numbers $\mathcal{I}_{ABC}=\mathcal{D}_{A}.%
\mathcal{D}_{B}.\mathcal{D}_{C}$ with $\mathcal{D}_{A}$ standing for the
seven $\left( D_{i},E_{a}\right) $. Multiplying formally both sides of eq(%
\ref{cyde}) by $E_{b}^{2}=E_{b}.E_{b}$ with $b=1,2,3$, we obtain the
following relationships between the triple intersection numbers,%
\begin{equation}
\sum_{i=1}^{4}\left( D_{i}.E_{b}^{2}\right) +\sum_{a=1}^{3}\left(
E_{a}.E_{b}^{2}\right) =0
\end{equation}%
These three relationships can be put into two convenient expressions; either
as
\begin{equation}
\sum_{i=1}^{4}\mathcal{J}_{i}^{b}+\sum_{a=1}^{3}\mathcal{I}_{a}^{b}=0
\label{JI}
\end{equation}%
where we have set $\mathcal{J}_{i}^{b}=D_{i}.E_{b}^{2}$ and $\mathcal{I}%
_{a}^{b}=E_{a}.E_{b}^{2}$; or into a more familiar form like
\begin{equation}
\sum_{A=1}^{7}q_{A}^{b}=0  \label{IJ}
\end{equation}%
with $q_{A}^{b}=\left( \mathcal{J}_{i}^{b},\mathcal{I}_{a}^{b}\right) $
standing three generalised Mori-vectors denoted below like $\mathbf{q}^{b}$ (%
$b=1,2,3$). The second expression is precisely the relation that we have in
gauged linear sigma model (GLSM) realisation of toric Calabi-Yau threefolds
\textrm{\cite{6b}}. Regarding the CY relation (\ref{IJ}), notice that it is
quite similar to the well known relation
\begin{equation}
\sum_{A=0}^{r}\left( Q_{A}^{b}\right) _{ADE}=0  \label{2d}
\end{equation}%
giving the CY condition we encounter in the study of complex 2d ADE surfaces
describing the resolution of ALE spaces with ADE singularities. In this
regards, recall that these complex ADE surfaces play a central role in the
geometric engineering of 4D $\mathcal{N}=2$ super QFTs from type IIA\ string
on Calabi-Yau threefolds given by ADE geometries fibered over the complex
line $\mathbb{C}$ \textrm{\cite{3b,4b,5f,3f,4f,6f}}. For these ADE
geometries which can be imagined in terms of orbifolds $\mathbb{C}^{2}$/$%
\Gamma $ with $\Gamma $ a discrete subgroup in SU$\left( 2\right) $, the
expression the (Mori-) vectors $\mathbf{Q}_{ADE}^{b}=\left( Q_{A}^{b}\right)
_{ADE}$ can be written down. For the example of the complex A$_{3}$ surface,
the three Mori- vectors read as follows%
\begin{equation}
\left( \mathbf{Q}^{a}\right) _{SU_{4}}=\left(
\begin{array}{ccccc}
1 & -2 & 1 & 0 & 0 \\
0 & 1 & -2 & 1 & 0 \\
0 & 0 & 1 & -2 & 1%
\end{array}%
\right)  \label{30}
\end{equation}%
where the Cartan matrix K$\left( SU_{4}\right) $ of the Lie algebra of the SU%
$\left( 4\right) $ gauge symmetry appears as a square sub-matrix of the
above $\left( \mathbf{Q}^{a}\right) _{SU_{4}}$. Recall that $K\left(
SU_{4}\right) $ is given by
\begin{equation}
K\left( SU_{4}\right) =-\left(
\begin{array}{ccc}
-2 & 1 & 0 \\
1 & -2 & 1 \\
0 & 1 & -2%
\end{array}%
\right)  \label{3a}
\end{equation}%
For the case of the CY graph $\mathcal{G}_{\hat{X}(Y^{4,0})}^{{\small SU}_{%
{\small 4}}}$ we are interested in this study, and depicted by the Figure
\textbf{\ref{y4-geo}},
the three generalised Mori- vectors $\left( \mathbf{q}^{a}\right) _{SU_{4}}$
are given by%
\begin{equation}
\left( \mathbf{q}^{a}\right) _{SU_{4}}=\left(
\begin{array}{ccccccc}
D_{1} & D_{2} & E_{1} & E_{2} & E_{3} & D_{3} & D_{4} \\
2 & 2 & -8 & 2 & 0 & 2 & 0 \\
2 & 0 & 2 & -8 & 2 & 2 & 0 \\
2 & 0 & 0 & 2 & -8 & 2 & 2%
\end{array}%
\right)  \label{qa}
\end{equation}%
This is a $3\times 7$ rectangular matrix that contains the square 3$\times $%
3 sub-matrix $\mathcal{I}_{b}^{a}$ \textrm{defined like} $E_{a}.E_{b}^{2}$
\textrm{and reading as follows}
\begin{equation}
\mathcal{I}\left( SU_{4}\right) _{a}^{b}=\left(
\begin{array}{ccc}
E_{1} & E_{2} & E_{3} \\
-8 & 2 & 0 \\
2 & -8 & 2 \\
0 & 2 & -8%
\end{array}%
\right)  \label{3i}
\end{equation}%
The diagonal terms $\mathcal{I}_{a}^{a}$ ($a=b$) describe precisely the
triple self intersections of the compact divisors namely $E_{a}^{3}=-8$. The
off diagonal terms $\mathcal{I}_{a}^{b}$ ($a\neq b$) describes the
intersection between the compact divisor $E_{a}$ and the compact curve $%
E_{b}^{2}.$ \newline
\textrm{As eqs(\ref{qa}) and (\ref{3i}) are one of the results of this
study, it is interesting to comment them by describing their content and
exploring their relationship with ADE Dynkin diagrams. These comments are as
listed below: }\newline
$\left( \mathbf{1}\right) $ First, recall that $E^{3}=-8$ is the triple self
intersection of the Hirzebruch surface $\mathbb{F}_{0}$ given by a complex
projective curve $\mathbb{P}^{1}$ fibered over another $\mathbb{P}^{1}$. So,
eq\textrm{(\ref{3i}) describes three }$\left( \mathbb{F}_{0}\right)
_{1},\left( \mathbb{F}_{0}\right) _{2}$ and $\left( \mathbb{F}_{0}\right)
_{3}$\textrm{\ intersecting transversally. The cross intersection described
by (\ref{3i}) concerns }$\left( \mathbb{F}_{0}\right) _{a}.C_{b}$ with $%
C_{b}=\left( \mathbb{F}_{0}\right) _{b}^{2}$ and $a\neq b$\textrm{. }\newline
$\left( \mathbf{2}\right) $ T\textrm{he quantity }$q_{A}^{b}=\mathcal{D}%
_{A}.E_{b}^{2}$ given by (\ref{fr}) describes the graph of the toric
Calabi-Yau threefolds $\hat{X}(Y^{4,0}).$ This quantity is interesting from
\textrm{various} views; in particular the three following: \newline
$\left( \mathbf{i}\right) $ The $q_{A}^{b}$ of the Calabi-Yau threefolds $%
\hat{X}(Y^{4,0})$ is the analogue of $Q_{A}^{b}$ eq(\ref{30}) associated
with Calabi-Yau twofolds (CY2). Then the $q_{A}^{b},$ concerning 4-cycles,
can be imagined as a generalisation of the Mori vector $Q_{A}^{b}$ dealing
with 2-cycles. As such $q_{A}^{b}$ and $Q_{A}^{b}$ can be put in
correspondence. This link is also supported by the fact that both $q_{A}^{b}$
and $Q_{A}^{b}$ are based on SU$\left( 4\right) $ and both obey the CY
condition namely $\sum q_{A}^{b}=0$ (\ref{IJ}) and $\sum Q_{A}^{b}=0$ (\ref%
{2d})$.$\newline
$\left( \mathbf{ii}\right) $ As for $q_{A}^{b}$ describing the toric $\hat{X}%
(Y^{4,0}),$ with graphic representation given by the Figure \textbf{\ref%
{y4-geo}}, the $Q_{A}^{b}$ describes also a toric CY2 surface $\hat{Z}%
_{SU_{4}}.$ This complex surface also has a graphic representation formally
similar to the vertical line of the Figure \textbf{\ref{y4-geo}}; that is
the line containing the red nodes.\ Recall that $\hat{Z}_{SU_{4}}$ is given
by the resolution of ALE space $\mathbb{C}^{2}/\mathbb{Z}_{4}$. The compact
part of the associated toric diagram is given by the Figure \textbf{\ref{A3}}%
-a where the nodes describe three intersecting $\mathbb{CP}^{1}$ curves.
\newline
$\left( \mathbf{iii}\right) $ The above comments done for SU$\left( 4\right)
$ holds in fact for the full SU$\left( p\right) $ family with $p\geq 2$. So,
the correspondence between $q_{A}^{b}$ and $Q_{A}^{b}$ is a general property
valid for SU$\left( p\right) _{0}$ gauge models in 5D. This correspondence
holds also for the intersection matrices $\mathcal{I}\left( SU_{p}\right)
_{a}^{b}$ and $K\left( SU_{p}\right) _{a}^{b}$ associated with the compact
parts in $q_{A}^{b}$ and $Q_{A}^{b}$ respectively. However, the graph of $%
K_{ab}$ is just the Dynkin diagram of the Lie algebra of SU$\left( 4\right) $%
. In this regards, recall that we have $K_{ab}=\alpha _{a}^{\vee }.\alpha
_{b}$ where the $\alpha _{a}$'s stand for the simple roots and the $\alpha
_{a}^{\vee }=2\alpha _{a}/\alpha _{a}^{2}$ for the co-roots. Clearly for SU$%
\left( p\right) $, we have $\alpha _{a}^{2}=2$. The Cartan matrix $K_{ab}$
has also an interpretation in terms of intersecting 2-cycles C$_{a}^{{\small %
(2)}}$ in the second homology group H$_{2}$; that is $C_{a}^{{\small (2)}%
}.C_{b}^{{\small (2)}}=-K_{ab}.$\newline
From this description, a natural question arises. Could the intersection
matrix \textrm{(\ref{3i}),} thought of as $\mathcal{I}_{ab}=E_{a}^{\vee
}.E_{b}$ with $E_{a}^{\vee }=E_{a}^{2},$ also has a similar algebraic
interpretation as $K_{ab}=\alpha _{a}^{\vee }.\alpha _{b}$? For example,
could $\mathcal{I}_{ab}$ be a generalized Cartan matrix $\mathcal{K}_{ab}^{%
{\small (gen)}}$\textrm{\  \cite{hyp}} or a cousin object of $\mathcal{K}%
_{ab}^{{\small (gen)}}$? In this regards, notice that like for $-K_{SU_{4}}$%
, the matrix $-\mathcal{I}_{SU_{4}}$ is an integer matrix with positive
entries on the diagonal and non positive off diagonal entries. At first
sight one might suspect this matrix to be a generalised Cartan matrix.
However, though it is symmetric and has a positive determinant ---$\det (-%
\mathcal{I}_{SU_{4}})>0$---, we have not found an algebraic interpretation
of this matrix. It is not either a generalised Cartan matrix of Borcherds
type. Progress in this direction will be reported in a future occasion.%
\newline
We end this section by noticing that eq(\ref{qa}) is a particular solution
of the Calabi-Yau condition (\ref{IJ}). It relies on the equality
\begin{equation}
E_{a}.E_{b}^{2}=E_{a}^{2}.E_{b}\qquad \Leftrightarrow \qquad E_{a}^{\vee
}.E_{b}=E_{b}^{\vee }.E_{a}
\end{equation}%
Other solutions of $\sum_{A}q_{A}^{b}=0$ violating the above symmetric
property can be also written down; they are omitted here.

\subsection{Leading members of the $\mathcal{G}_{\hat{X}(Y^{p,0})}^{%
{\protect \small SU}_{{\protect \small p}}}$ family}

\begin{figure}[tbph]
\begin{center}
\includegraphics[width=12cm]{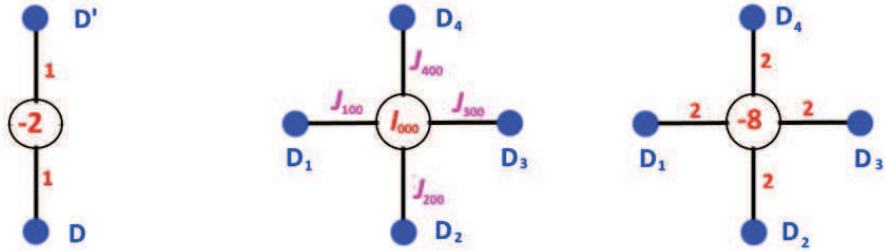}
\end{center}
\par
\vspace{-0.5cm}
\caption{On the left, the toric diagram of the $A_{1}$ geometry. It has one
compact 2-sphere, with self intersection $\left( -2\right) $, intersecting
two transverse non compact (blue) 2-spheres once. On the right, the graph $%
\mathcal{G}_{\hat{X}(Y^{2,0})}^{{\protect \small SU}_{{\protect \small 2}}}$
geometry having one compact 4-cycle, with triple self intersection $\mathcal{%
I}_{000}=-8$, intersecting four non compact (blue) 4-cycles. The Calabi-Yau
threefolds condition is ensured by the vanishing sum of the total charge.}
\label{F0-geo}
\end{figure}

\begin{figure}[tbph]
\begin{center}
\includegraphics[width=6cm]{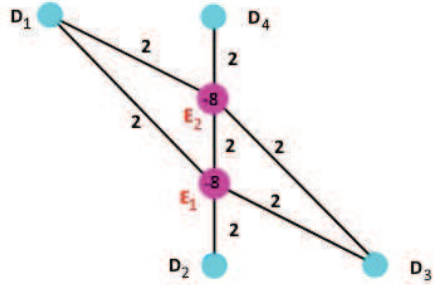}
\end{center}
\par
\vspace{-0.5cm}
\caption{The CY graph $\mathcal{G}_{\hat{X}(Y^{3,0})}^{{\protect \small SU}_{%
{\protect \small 3}}}$ of the toric threefolds $\hat{X}\left( Y^{3,0}\right) $
exhibiting manifestly the Calabi-Yau condition at each internal point of the
graph.}
\label{y30}
\end{figure}

In the above subsection, we have focussed on the CY graph $\mathcal{G}_{\hat{%
X}(Y^{4,0})}^{{\small SU}_{{\small 4}}}$ given by the Figure \textbf{\ref%
{y4-geo} }which is\textbf{\ }based on exhibiting the triple intersection
numbers of the compact divisors $E_{1},E_{2},E_{2}$ amongst themselves and
with the non compact $D_{1},D_{2},D_{3},D_{4}$. This CY graph is however the
\textrm{third} member of the family $\mathcal{G}_{\hat{X}(Y^{p,0})}^{{\small %
SU}_{{\small p}}}$ with $p\geq 2.$ Below, we give comments on the two first
CY graphs of this family.\newline
The first member of the $\mathcal{G}_{\hat{X}(Y^{p,0})}^{{\small SU}_{%
{\small p}}}$ family is given by $\mathcal{G}_{\hat{X}(Y^{2,0})}^{{\small SU}%
_{{\small 2}}}$ (p=2). It has four (external) non compact divisors $%
D_{1},D_{2},D_{3},D_{4}$; but only one internal compact divisor that we
denote $E_{0}$. So, there is one generalised Mori- vector given by%
\begin{equation}
\left( \mathbf{q}\right) _{{\small SU}_{{\small 2}}}=\left(
\begin{array}{ccccc}
{\small D}_{{\small 1}} & {\small D}_{{\small 2}} & {\small E}_{{\small 0}}
& {\small D}_{{\small 3}} & {\small D}_{{\small 4}} \\
2 & 2 & -8 & 2 & 2%
\end{array}%
\right)
\end{equation}%
where the CY condition, given by the vanishing of the trace of $\left(
\mathbf{q}\right) _{{\small SU}_{{\small 2}}},$ is manifestly exhibited. The
diagram representing the CY graph $\mathcal{G}_{\hat{X}(Y^{2,0})}^{{\small SU%
}_{{\small 2}}}$ is given by the picture on the right side of the Figure
\textbf{\ref{F0-geo}}. On the left side of this figure, we have given the
picture of the standard $A_{1}$ geometry of ALE space involving complex
projective curves with self intersection $-2$.\textbf{\ }
Notice that the Calabi-Yau threefold $\hat{X}(Y^{2,0})$ is precisely $\hat{X}%
(\mathbb{F}_{0})$, the toric threefold based on the Hirzebruch surface $%
\mathbb{F}_{0}$ which is known to have a triple self intersection $\left(
-8\right) $. Notice also that this graph has outer-automorphisms given by
the mirror $\left( \mathbb{Z}_{2}^{x}\right) _{\Delta ^{{\small SU}%
_{2}}}\times \left( \mathbb{Z}_{2}^{y}\right) _{\Delta ^{{\small SU}_{2}}}$
fixing E$_{0}$ and acting by the exchange $D_{1}\leftrightarrow D_{3}$ and $%
D_{2}\leftrightarrow D_{4}$.\newline
Concerning the second member of the family namely $\mathcal{G}_{\hat{X}%
(Y^{P,0})}^{{\small SU}_{{\small P}}}$ with $p=3$, it has four external non
compact divisors $D_{1},D_{2},D_{3},D_{4}$; but two compact divisors E$_{1}$
and E$_{2}$. For this case, there are two generalised Mori- vectors given by%
\begin{equation}
\left( \mathbf{q}^{a}\right) _{{\small SU}_{{\small 3}}}=\left(
\begin{array}{cccccc}
D_{1} & D_{2} & E_{1} & E_{2} & D_{3} & D_{4} \\
2 & 2 & -8 & 2 & 2 & 0 \\
2 & 0 & 2 & -8 & 2 & 2%
\end{array}%
\right)
\end{equation}%
The representative CY graph $\mathcal{G}_{\hat{X}(Y^{3,0})}^{{\small SU}_{%
{\small 3}}}$ is depicted by the Figure \textbf{\ref{y30}}.
Notice that the graph $\mathcal{G}_{\hat{X}(Y^{3,0})}^{{\small SU}_{{\small 3%
}}}$ has an outer-automorphism symmetry group $H_{\Delta ^{{\small SU}%
_{3}}}^{{\small outer}}$ given by $\left( \mathbb{Z}_{2}^{x}\right) _{\Delta
^{{\small SU}_{3}}}\times \left( \mathbb{Z}_{2}^{y}\right) _{\Delta ^{%
{\small SU}_{3}}}$; but with no fix divisor. The full $H_{\Delta ^{{\small SU%
}_{3}}}^{{\small outer}}$ acts by the exchange $E_{1}\leftrightarrow E_{2}$,
$D_{1}\leftrightarrow D_{3}$ and $D_{2}\leftrightarrow D_{4}$. This $%
H_{\Delta ^{{\small SU}_{3}}}^{{\small outer}}$ is a subsymmetry of $\mathbb{%
Z}_{6}$. It is generated by the product of three transpositions namely $\tau
\circ \tau ^{\prime }\circ \tau ^{\prime \prime }$ with transpositions given
by $\tau =\left( E_{1}E_{2}\right) ,$ $\tau ^{\prime }=\left(
D_{1}D_{3}\right) $ and $\tau ^{\prime \prime }=\left( D_{2}D_{4}\right) $.

\section{Symplectic graphs and quivers}

In this section, we first build the symplectic CY graph $\mathcal{G}_{\hat{X}%
\left( Y^{4,0}\right) }^{{\small SP}_{4}}$ by starting from the unitary $%
\mathcal{G}_{\hat{X}\left( Y^{4,0}\right) }^{{\small SU}_{4}}$ and using
folding ideas under $\left( \mathbb{Z}_{2}^{x}\right) _{\Delta ^{{\small SU}%
_{4}}}\times \left( \mathbb{Z}_{2}^{y}\right) _{\Delta ^{{\small SU}_{4}}}$.
Then, we construct the symplectic quiver $Q_{\hat{X}\left( Y^{4,0}\right) }^{%
{\small SP}_{4}}$ with symplectic SP$\left( 4,\mathbb{R}\right) $ gauge
symmetry by using the unitary BPS quiver $Q_{\hat{X}\left( Y^{4,0}\right) }^{%
{\small SU}_{4}}$ and outer-automorphisms $\left( \mathbb{Z}%
_{2}^{outer}\right) _{Q^{{\small SU}_{4}}}$.

\subsection{Symplectic CY graph $\mathcal{G}_{\hat{X}\left( Y^{4,0}\right)
}^{{\protect \small SP}_{{\protect \small 4}}}$}

We start by the toric data of $\Delta _{\hat{X}\left( Y^{4,0}\right) }^{%
{\small SU}_{4}}$ given by Table \textbf{\ref{q}}. Because these data are
defined up to a global shift; we translate the points of $\Delta _{\hat{X}%
\left( Y^{4,0}\right) }^{{\small SU}_{4}}$ by $\left( 0,-2\right) $. So the
values of the $\mathbf{w}_{i}$ and $\mathbf{\upsilon }_{a}$ points ---Table
\textbf{\ref{q}}--- of the previous toric diagram gets mapped to new points
that we present as follows
\begin{table}[h]
\centering \renewcommand{\arraystretch}{1.2} $%
\begin{tabular}{|l||l|l|l|l|l||l|l||}
\hline
$\hat{X}(Y^{4,0})$ & \multicolumn{5}{||l}{\  \  \  \  \  \  \  \  \  \  \  \  \  \  \  \  \ A%
$_{3}$- geometry} & \multicolumn{2}{||l||}{transverse geometry} \\ \hline
$\  \Delta _{\hat{X}\left( Y^{4,0}\right) }^{SU_{4}}$ & $\  \  \upsilon _{-2}$
& $\  \  \upsilon _{-1}$ & $\  \  \upsilon _{0}$ & $\  \  \upsilon _{+1}$ & $\  \
\upsilon _{+2}$ & $\  \  \ w_{-1}$ & $\  \ w_{+1}$ \\ \hline \hline
\ points & $\left( 0,-2\right) $ & $\left( 0,-1\right) $ & $\left(
0,0\right) $ & $\left( 0,+1\right) $ & $\left( 0,+2\right) $ & $\left(
-1,+2\right) $ & $\left( +1,-2\right) $ \\ \hline \hline
divisors & \ E$_{-2}$ & \ E$_{-1}$ & \ E$_{0}$ & \ E$_{+1}$ & \ E$_{+2}$ & \
D$_{-1}$ & \ D$_{+1}$ \\ \hline \hline
\end{tabular}%
$%
\caption{Toric data exhibiting manifestly outer-automorphism symmetry{}}
\label{t}
\end{table}
where we have set $D_{-2}\equiv E_{-2}$ and $D_{+2}\equiv E_{+2}$. With this
parametrisation, the internal point $\mathbf{\upsilon }_{0}=\left(
0,0\right) $ is at the centre of the toric diagram. Moreover, the toric $%
\Delta _{\hat{X}\left( Y^{4,0}\right) }^{{\small SU}_{4}}$ is invariant
under the outer-automorphism symmetry group $H_{\Delta ^{{\small SU}%
_{4}}}^{outer}\simeq \left( \mathbb{Z}_{2}^{x}\right) _{\Delta ^{{\small SU}%
_{4}}}\times \left( \mathbb{Z}_{2}^{y}\right) _{\Delta ^{{\small SU}_{4}}}$
mapping the points $\mathbf{w}_{\pm i},\mathbf{\upsilon }_{0},\mathbf{%
\upsilon }_{\pm a}$ into the symmetric ones namely $\mathbf{w}_{\mp i},%
\mathbf{\upsilon }_{0},\mathbf{\upsilon }_{\mp a}.$ Because of the property $%
\mathbf{w}_{\mp i}=-\mathbf{w}_{\pm i}$ and $\mathbf{\upsilon }_{\mp a}=-%
\mathbf{\upsilon }_{\pm a},$ the outer-automorphism $H_{\Delta ^{{\small SU}%
_{4}}}^{outer}$ acts as a parity symmetry of the toric diagram,
\begin{equation}
H_{\Delta ^{{\small SU}_{{\small 4}}}}^{outer}:\left( \mathbf{w}_{\mp i},%
\mathbf{0},\mathbf{\upsilon }_{\mp a}\right) \qquad \rightarrow \qquad
\left( -\mathbf{w}_{\pm i},\mathbf{0},-\mathbf{\upsilon }_{\pm a}\right)
\label{g0}
\end{equation}%
Notice that the outer-automorphism parity $H_{\Delta ^{{\small SU}%
_{4}}}^{outer}$ is isomorphic to the group product $\left( \mathbb{Z}%
_{2}^{x}\right) _{\Delta ^{{\small SU}_{4}}}\times \left( \mathbb{Z}%
_{2}^{y}\right) _{\Delta ^{{\small SU}_{4}}}$ generated by the reflections
in x- and y- directions acting as follows
\begin{equation}
\begin{tabular}{lllll}
$\left( \mathbb{Z}_{2}^{x}\right) _{\Delta ^{{\small SU}_{4}}}$ & $:$ & $%
\left( n_{x},n_{y}\right) $ & $\rightarrow $ & $\left( -n_{x},n_{y}\right) $
\\
$\left( \mathbb{Z}_{2}^{y}\right) _{\Delta ^{{\small SU}_{4}}}$ & $:$ & $%
\left( n_{x},n_{y}\right) $ & $\rightarrow $ & $\left( n_{x},-n_{y}\right) $
\\
$\left( \mathbb{Z}_{2}^{x}\right) _{\Delta ^{{\small SU}_{4}}}\times \left(
\mathbb{Z}_{2}^{y}\right) _{\Delta ^{{\small SU}_{4}}}$ & $:$ & $\left(
n_{x},n_{y}\right) $ & $\rightarrow $ & $\left( -n_{x},-n_{y}\right) $%
\end{tabular}%
\end{equation}%
where the $\left( n_{x},n_{y}\right) $'s stand for the values of the
external and the internal points of the toric diagram. So, the triangulated $%
\Delta _{\hat{X}\left( Y^{2r,0}\right) }^{{\small SU}_{{\small 2r}}}$ is
invariant under the outer-automorphism symmetry group with the central point
$\mathbf{\upsilon }_{0}$ being the unique fix point of $H_{\Delta ^{{\small %
SU}_{4}}}^{outer}.$ \newline
By folding the CY graph $\mathcal{G}_{\hat{X}\left( Y^{4,0}\right) }^{%
{\small SU}_{4}}$ under the parity symmetry $\left( \mathbb{Z}%
_{2}^{x}\right) _{\Delta ^{{\small SU}_{4}}}\times \left( \mathbb{Z}%
_{2}^{y}\right) _{\Delta ^{{\small SU}_{4}}}$, we end up with a new CY graph
\begin{equation}
\mathcal{\tilde{G}}_{\hat{X}\left( Y^{4,0}\right) }^{{\small SP}_{4}}=\frac{%
\mathcal{G}_{\hat{X}\left( Y^{4,0}\right) }^{{\small SU}_{{\small 4}}}}{%
\left( \mathbb{Z}_{2}^{x}\right) _{\Delta ^{{\small SU}_{4}}}\times \left(
\mathbb{Z}_{2}^{y}\right) _{\Delta ^{{\small SU}_{4}}}}
\end{equation}%
having $2+2=4$ points given, up to identifications, by $\mathbf{w}%
_{-1}\equiv \mathbf{w}_{+1}$, $\mathbf{w}_{-2}\equiv \mathbf{w}_{+2};$ and $%
\mathbf{\upsilon }_{0}$ as well as $\mathbf{\upsilon }_{-1}\equiv \mathbf{%
\upsilon }_{+1}.$ The CY graph $\mathcal{\tilde{G}}_{\hat{X}\left(
Y^{4,0}\right) }^{{\small SP}_{4}}$ is depicted by the Figure \textbf{\ref%
{g40-c}}.
\begin{figure}[tbph]
\begin{center}
\includegraphics[width=6cm]{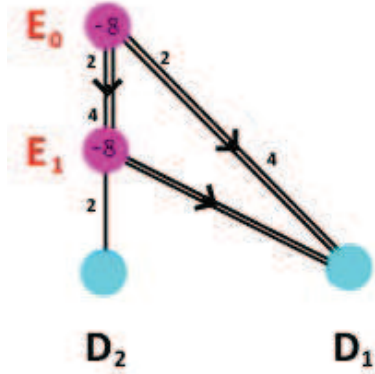}
\end{center}
\par
\vspace{-0.5cm}
\caption{The folding CY graphs $\mathcal{G}_{\hat{X}\left( Y^{2r,0}\right)
}^{{\protect \small SU}_{{\protect \small 4}}}$ under parity symmetry $\left(
\mathbb{Z}_{2}^{x}\right) _{\Delta ^{{\protect \small SU}_{4}}}\times \left(
\mathbb{Z}_{2}^{y}\right) _{\Delta ^{{\protect \small SU}_{4}}}$. On the
vertical line, one has a vertical chain that looks like Dynkin diagram of
the C$_{2}\simeq sp\left( 4,\mathbb{R}\right) $ Lie algebras. Here, we have $%
E_{0}^{2}.D_{1}=E_{1}^{2}.D_{1}=4$ and $E_{0}^{2}.E_{1}\neq E_{1}^{2}.E_{0}$%
. }
\label{g40-c}
\end{figure}
The generalised Mori- vectors $\mathbf{\tilde{q}}^{1}$ and $\mathbf{\tilde{q}%
}^{2}$ associated with the symplectic CY graph $\mathcal{\tilde{G}}_{\hat{X}%
\left( Y^{4,0}\right) }^{{\small SP}_{4}}$ have each four components $\tilde{%
q}_{\beta }^{a}$. They are given by%
\begin{equation}
\tilde{q}_{\beta }^{a}=\left(
\begin{array}{cccc}
{\small E}_{\pm {\small 2}} & {\small E}_{\pm {\small 1}} & {\small E}_{%
{\small 0}} & {\small D}_{\pm {\small 1}} \\
2 & -8 & 2 & 4 \\
0 & 4 & -8 & 4%
\end{array}%
\right)
\end{equation}%
The $2\times 2$ square submatrix of the above rectangular $\tilde{q}_{\beta
}^{a},$ associated with the triple intersections of the compact divisors $%
E_{0}$ and $E_{\pm {\small 1}},$ is given by%
\begin{equation}
\mathcal{I}_{b}^{a}=\left(
\begin{array}{cc}
-8 & 2 \\
4 & -8%
\end{array}%
\right)  \label{2i}
\end{equation}%
Remarkably, this intersection matrix $\mathcal{I}_{b}^{a}=E_{a}^{2}.E_{b}$
between the compact divisors is non symmetric. It can be put in
correspondence with the non symmetric Cartan matrix $K\left( C_{2}\right) $
of the symplectic C$_{2}$ Lie algebra given by%
\begin{equation}
K\left( C_{2}\right) =\left(
\begin{array}{cc}
2 & -1 \\
-2 & 2%
\end{array}%
\right)  \label{2a}
\end{equation}%
The construction we have done for the particular $\mathcal{\tilde{G}}_{\hat{X%
}\left( Y^{4,0}\right) }^{{\small SP}_{4}}$ can be straightforwardly
generalized to $\mathcal{\tilde{G}}_{\hat{X}\left( Y^{2r,0}\right) }^{%
{\small SP}_{2r}}$ with $r\geq 2.$ The generic intersection matrix $\mathcal{%
I}_{ab}$ with a,b=1, ....r; has a quite similar form as (\ref{2i}). It is
non symmetric $E_{a}^{2}.E_{b}\neq E_{b}^{2}.E_{a}$ and can be put in
correspondence with the Cartan matrix of the symplectic Lie algebra C$_{r};$
\textrm{see the discussion given after} eq\textrm{(\ref{3i})}.\newline
We end this subsection by making a comment on the folding of the family of
Calabi-Yau graphs $\mathcal{G}_{\hat{X}\left( Y^{p,0}\right) }^{{\small SU}_{%
{\small p-1}}}$ with respect to the factor $\left( \mathbb{Z}_{2}^{x}\right)
_{\Delta ^{{\small SU}_{p}}}$. It may be imagined as a partial folding in
the transverse geometry represented by the points $\mathbf{w}_{1}$ and $%
\mathbf{w}_{3}$ of the toric diagram as indicated by Table \textbf{\ref{q}}.
Recall that $\left( \mathbb{Z}_{2}^{x}\right) _{\Delta ^{{\small SU}_{p}}}$
fixes all internal points $\upsilon _{a}$ of the toric diagrams $\Delta _{%
\hat{X}\left( Y^{p,0}\right) }^{{\small SU}_{{\small p-1}}}$ as well as the
two external $\mathbf{w}_{2}$ and $\mathbf{w}_{4}$; but exchanges the two
other external $\mathbf{w}_{1}$ and $\mathbf{w}_{3}$. The folded
\begin{equation}
\mathcal{G}_{\hat{X}\left( Y^{p,0}\right) }^{{\small SU}_{{\small p-1}%
}}/\left( \mathbb{Z}_{2}^{x}\right) _{\Delta ^{{\small SU}_{p}}}
\end{equation}%
gives an exotic Calabi-Yau diagram; which for the example p=3 is given by
the Figure \textbf{\ref{3g}}.
\begin{figure}[tbph]
\begin{center}
\includegraphics[width=6cm]{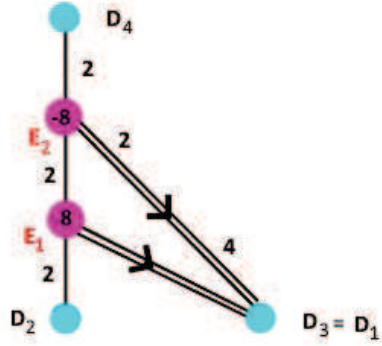}
\end{center}
\par
\vspace{-0.5cm}
\caption{The folding CY graphs $\mathcal{G}_{\hat{X}\left( Y^{3,0}\right) }^{%
{\protect \small SU}_{{\protect \small 3}}}$ under the partial
outer-automorphism symmetry group $\left( \mathbb{Z}_{2}^{x}\right) _{\Delta
^{{\protect \small SU}_{3}}}$. Because of the folding, the two divisors $%
D_{1} $ and $D_{3}$ merge. Here, we have $E_{1}^{2}.D_{1}=E_{2}^{2}.D_{1}=4$
and $E_{2}^{2}.E_{1}=E_{1}^{2}.E_{2}$.}
\label{3g}
\end{figure}
For this exotic folding, there are still two generalised Mori vectors that
are associated with the compact divisors E$_{1}$ and E$_{2}$. These vectors
are given by
\begin{equation}
\mathbf{\tilde{q}}^{a}=\left(
\begin{array}{ccccc}
{\small D}_{{\small 2}} & {\small E}_{{\small 1}} & {\small E}_{{\small 2}}
& {\small D}_{3} & {\small D}_{4} \\
2 & -8 & 2 & 4 & 0 \\
0 & 2 & -8 & 4 & 2%
\end{array}%
\right)
\end{equation}%
The intersection matrix $\mathcal{I}_{ab}=E_{a}^{2}.E_{b}$ concerning the
compact divisors is given by
\begin{equation}
\mathcal{I}_{ab}=\left(
\begin{array}{cc}
-8 & 2 \\
2 & -8%
\end{array}%
\right)
\end{equation}%
it is symmetric as in the unitary case.

\subsection{BPS quiver with SP$\left( 4,\mathbb{R}\right) $ invariance}

\begin{figure}[tbph]
\begin{center}
\includegraphics[width=8cm]{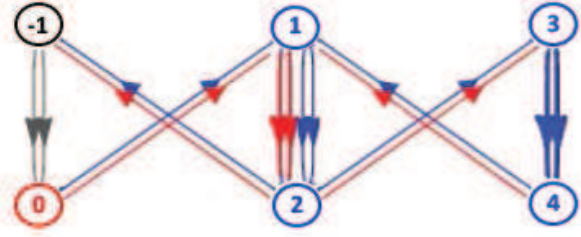}
\end{center}
\par
\vspace{-0.5cm}
\caption{The BPS quiver $Q_{{\protect \small \hat{X}(Y}^{{\protect \small 4,0}}%
{\protect \small )}}^{{\protect \small SP}_{{\protect \small 4}}}$ resulting by
the folding of the $Q_{{\protect \small \hat{X}(Y}^{{\protect \small 4,0}}%
{\protect \small )}}^{{\protect \small SU}_{{\protect \small 4}}}$ under the
mirror $\left( \mathbb{Z}_{2}^{{\protect \small outer}}\right) _{Q}.$ The
nodes $\left \{ 1,2,3,4\right \} $ denote the elementary BPS particles
associated with the rank of the C$_{2}$ Lie algebra.}
\label{sp4r}
\end{figure}

The BPS quiver $Q_{\hat{X}\left( Y^{4,0}\right) }^{{\small SP}_{{\small 4}}}$
with SP$\left( 4,\mathbb{R}\right) $ gauge invariance is obtained by folding
the unitary $Q_{\hat{X}\left( Y^{4,0}\right) }^{{\small SU}_{{\small 4}}}$
by its outer-automorphism symmetry group $H_{{\small Q}^{{\small SU}_{%
{\small 4}}}}^{{\small outer}}$ whose action on the quiver nodes and the
arrows is constructed below. The BPS quiver $Q_{{\small \hat{X}(Y}^{{\small %
4,0}})}^{{\small SU}_{{\small 4}}}$ has 8 nodes $\left \{ j\right \} $ with $%
j=1,...,8$; and 16 oriented arrows $\left \langle j|l\right \rangle $ as
depicted by the Figure \textbf{\ref{3}.} Clearly, the BPS quiver $Q_{{\small
\hat{X}(Y}^{{\small 4,0}})}^{{\small SU}_{{\small 4}}}$ has a non trivial
outer- automorphism symmetry group with two factors given by
\begin{equation}
H_{{\small Q}^{{\small SU}_{{\small 4}}}}^{{\small outer}}=\left( \mathbb{Z}%
_{4}\right) _{Q^{{\small SU}_{{\small 4}}}}\times \left( \mathbb{Z}_{2}^{%
{\small outer}}\right) _{Q^{{\small SU}_{{\small 4}}}}  \label{4z}
\end{equation}%
The factor $\left( \mathbb{Z}_{4}\right) _{Q^{{\small SU}_{{\small 4}}}}$
has no fix quiver-node and no fix quiver-arrows; while $\left( \mathbb{Z}%
_{2}^{{\small outer}}\right) _{Q^{{\small SU}_{{\small 4}}}}$ has fix nodes
and arrows. The transformations under $\left( \mathbb{Z}_{4}\right) _{Q^{%
{\small SU}_{{\small 4}}}}$ are given eq(\ref{z4}); and the change under $%
\left( \mathbb{Z}_{2}^{{\small outer}}\right) _{Q^{{\small SU}_{{\small 4}%
}}} $ is as in eq(\ref{z2}). Thus, it is the symmetry $\left( \mathbb{Z}%
_{2}^{{\small outer}}\right) _{Q^{{\small SU}_{{\small 4}}}}$ that is
important in our folding construction as it has fix nodes and arrows.
\newline
For the transformations under $\left( \mathbb{Z}_{4}\right) _{Q^{{\small SU}%
_{{\small 4}}}}$, the nodes are transformed as follows%
\begin{equation}
\left( \mathbb{Z}_{4}\right) _{Q^{{\small SU}_{{\small 4}}}}:%
\begin{tabular}{lll}
$\left \{ 2a-1\right \} $ & $\rightarrow $ & $\left \{ 2a+7\right \} $ \\
$\  \  \  \  \  \left \{ 2a\right \} $ & $\rightarrow $ & $\left \{ 2a+8\right \}
$%
\end{tabular}%
\end{equation}%
\textrm{This change indicates} that the nodes $\left \{ 7\right \} $ and $%
\left \{ 8\right \} $ can be also denoted like $\left \{ \bar{1}\right \} $
and $\left \{ 0\right \} $ respectively where the label $\bar{1}$ refers to $%
-1\equiv 7$ and $0\equiv 8$. Regarding the action of the $\left( \mathbb{Z}%
_{2}^{{\small outer}}\right) _{Q^{{\small SU}_{{\small 4}}}}$ symmetry, we
have the following transformations of the nodes%
\begin{equation}
\left( \mathbb{Z}_{2}^{{\small outer}}\right) _{Q^{{\small SU}_{{\small 4}%
}}}:%
\begin{tabular}{lll}
$\left \{ 2a-1\right \} $ & $\rightarrow $ & $\left \{ 7-2a\right \} $ \\
$\  \  \  \  \  \left \{ 2a\right \} $ & $\rightarrow $ & $\left \{ 8-2a\right \}
$%
\end{tabular}%
\end{equation}%
showing that four quiver- nodes amongst the eight ones are fixed. They
concern the pair $\left \{ 3\right \} ,\left \{ 4\right \} $ and the pair $%
\left \{ 7\right \} ,$ $\left \{ 8\right \} $.\newline
By folding $Q_{{\small \hat{X}(Y}^{{\small 4,0}})}^{{\small SU}_{{\small 4}%
}} $ with respect to $\left( \mathbb{Z}_{2}^{{\small outer}}\right) _{Q^{%
{\small SU}_{{\small 4}}}}$, we obtain a BPS\ quiver interpreted as the BPS
quiver $Q_{{\small \hat{X}(Y}^{{\small 4,0}})}^{{\small SP}_{{\small 4}}}$
with an SP$\left( 4,\mathbb{R}\right) $ gauge symmetry. This folded BPS
quiver has 6 nodes namely $\left \{ 1,2,3,4,7,8\right \} $ (the old $%
\left
\{ 5\right \} $ and $\left \{ 6\right \} $ omitted due to folding).
This new set of nodes can be also denoted as $\left \{ \bar{1}%
,0,1,2,3,4\right \} $ where we have renamed the Kronecker quiver $\left \{
7,8\right \} $ like $\left \{ \bar{1},0\right \} $. The resulting BPS quiver
$Q_{{\small \hat{X}(Y}^{{\small 4,0}})}^{{\small SP}_{{\small 4}}}$ is as
depicted in the Figure \textbf{\ref{sp4r}}.
In addition to the nodes, the folded BPS quiver has 16 oriented arrows
distributed as in the Table \textbf{\ref{5}}
\begin{table}[h]
\centering \renewcommand{\arraystretch}{1.2} $%
\begin{tabular}{|l|l|l|l|l|l|l|l|}
\hline
\  \ arrows & $\left \langle 1|2\right \rangle ^{\alpha a}$ & $\left \langle
3|4\right \rangle ^{\alpha }$ & $\left \langle 7|8\right \rangle ^{\alpha }$
& $\left \langle 2|3\right \rangle ^{a}$ & $\left \langle 8|1\right \rangle
^{a}$ & $\left \langle 4|1\right \rangle ^{a}$ & $\left \langle 2|7\right
\rangle ^{a}$ \\ \hline
\  \ matter & $X_{12}^{\alpha a}$ & $X_{34}^{\alpha }$ & $X_{78}^{\alpha }$ &
$Z_{23}^{a}$ & $Z_{81}^{a}$ & $Y_{41}^{a}$ & $Y_{27}^{a}$ \\ \hline
multiplicity & $\  \  \ 4$ & $\  \  \ 2$ & $\  \  \ 2$ & $\  \  \ 2$ & $\  \  \ 2$ & $%
\  \  \ 2$ & $\  \ 2$ \\ \hline
\end{tabular}%
$%
\caption{Matter content of the BPS quiver $Q_{{\protect \small \hat{X}(Y}^{%
{\protect \small 4,0}})}^{{\protect \small SP}_{{\protect \small 4}}}$.}
\label{5}
\end{table}
where the complex $X_{12}^{\alpha a}$ with $\alpha =1,2$ and $a=1,2$ form a
quartet; and where the $U_{\ast }^{\alpha }$ and the $U_{\ast }^{a}$ are
doublets with $U$ standing for $X,Y$ and $Z$. With these complex
superfields, one can write down the SQM superpotential of the theory; it
will not be discussed here.

\section{Conclusion and comments}

In this paper, we have developed a method to construct a new family of 5D $%
\mathcal{N}=1$ supersymmetric QFT models compactified on \textrm{a circle}
with finite radius. This family of gauge models has symplectic SP$\left( 2r,%
\mathbb{R}\right) $ gauge invariance and is embedded in M-theory on CY3s
based on Sasaki-Einstein manifolds Y$^{p,0}.$ Recall that gauge models
engineered from M-theory on $\hat{X}\left( Y^{p,q}\right) $ are well known;
and they have unitary symmetries. So, our construction can be viewed as
widening the family of unitary models based on $\hat{X}\left( Y^{p,q}\right)
$ to include\emph{\ }the family of\emph{\ }symplectic invariant models. To
engineer this new theory, we started from the 5D $\mathcal{N}=1$ super QFT,
with unitary SU$\left( 4\right) $ gauge symmetry (corresponding to 2r=4),
embedded in M-theory compactification on the toric Calabi-Yau threefold $%
\hat{X}\left( Y^{4,0}\right) .$ \textrm{T}his complex 3d variety is a
resolution of a conical singularity based on the Sasaki-Einstein manifold $%
Y^{4,0}$. Then, we \textrm{have proposed a graph }to represent $\hat{X}%
\left( Y^{4,0}\right) $ by using the numbers $\mathcal{J}_{iab}$ and $%
\mathcal{I}_{abc}$ given by triple intersections of the 7 divisors of $\hat{X%
}\left( Y^{4,0}\right) $; four non compact D$_{i}$ and three compact E$_{a}$%
. This new graph, denoted as $\mathcal{G}_{\hat{X}\left( Y^{4,0}\right) }^{%
{\small SU}_{{\small 4}}}$, is given by a generalisation of the Mori-vectors
of the ADE geometries of ALE spaces. \textrm{It} is defined by eq(\ref{fr})
and, to our knowledge, it has not been used before. \textrm{We qualified the
graph }$\mathcal{G}_{\hat{X}\left( Y^{4,0}\right) }^{{\small SU}_{{\small 4}%
}}$\emph{\ }as a unitary CY graph, first because of the unitary $SU\left(
4\right) $ symmetry of the gauge fiber within $\hat{X}\left( Y^{4,0}\right) $%
; and second to distinguish it from the CY graph $\mathcal{G}_{\hat{X}\left(
Y^{4,0}\right) }^{{\small SP}_{{\small 4}}}$ having a symplectic $SP\left( 4,%
\mathbb{R}\right) $ gauge symmetry. The use of $\mathcal{G}_{\hat{X}\left(
Y^{4,0}\right) }^{{\small SU}_{{\small 4}}}$ has the merit to $\left(
\mathbf{i}\right) $ highlight the CY condition of the toric $\hat{X}\left(
Y^{4,0}\right) $; $\left( \mathbf{ii}\right) $ extend the usual complex A$%
_{3}$ surface describing the resolution of an ALE space with an SU$\left(
4\right) $ singularity; and $\left( \mathbf{iii}\right) $ to study non
trivial outer-automorphisms $H_{\Delta ^{{\small SU}_{{\small 4}}}}^{{\small %
outer}}$ of the toric diagram $\Delta _{\hat{X}\left( Y^{4,0}\right) }^{%
{\small SU}_{{\small 4}}}$. The outer-automorphism group $H_{\Delta ^{%
{\small SU}_{{\small 4}}}}^{{\small outer}}$ has a fixed internal point (a
compact divisor); and is used to build the symplectic CY graph $\mathcal{G}_{%
\hat{X}\left( Y^{4,0}\right) }^{{\small SP}_{{\small 4}}}$ by using the
folding $\mathcal{G}_{\hat{X}\left( Y^{4,0}\right) }^{{\small SU}_{{\small 4}%
}}/H_{\Delta ^{{\small SU}_{{\small 4}}}}^{{\small outer}}.$ After having
set the basis for the CY graphs to represent the toric threefolds $\hat{X}%
\left( Y^{p,q}\right) $, we turned to investigating the BPS particles by
constructing the symplectic BPS quiver $Q_{\hat{X}\left( Y^{4,0}\right) }^{%
{\small SP}_{{\small 4}}}$ that is associated with the symplectic CY graph $%
\mathcal{G}_{\hat{X}\left( Y^{4,0}\right) }^{{\small SP}_{{\small 4}}}$.
This BPS quiver is obtained by folding the unitary BPS $Q_{\hat{X}\left(
Y^{4,0}\right) }^{{\small SU}_{{\small 4}}}$ with respect to
outer-automorphisms $\left( \mathbb{Z}_{2}^{{\small outer}}\right) _{Q^{%
{\small SU}_{{\small 4}}}}$. Recall that the $Q_{\hat{X}\left(
Y^{4,0}\right) }^{{\small SU}_{{\small 4}}}$ has 8 nodes and 16 oriented
arrows respectively describing 8 elementary BPS particles and 16 chiral
superfields in SQM. The mirror symmetry $\left( \mathbb{Z}_{2}^{{\small outer%
}}\right) _{Q^{{\small SU}_{{\small 4}}}}$ fixes four nodes of $Q_{\hat{X}%
\left( Y^{4,0}\right) }^{{\small SU}_{{\small 4}}}$ and exchanges the four
others. It fixes four arrows and exchanges the 12 others.\newline
We end this conclusion by making two more comments regarding extensions of
the analysis done in this paper. \newline
The first extension concerns the building of symplectic BPS quivers $Q_{\hat{%
X}\left( Y^{2r,0}\right) }^{{\small SP}_{{\small 2r}}}$ with generic rank.
This is achieved by starting from the unitary quiver $Q_{\hat{X}\left(
Y^{2r,0}\right) }^{{\small SU}_{{\small 2r}}}$ with rank 2r-1 and use
folding ideas. The resulting symplectic quivers $Q_{\hat{X}\left(
Y^{2r,0}\right) }^{{\small SP}_{{\small 2r}}}$ are associated with the toric
threefolds obtained by folding the unitary $Q_{\hat{X}\left( Y^{2r,0}\right)
}^{{\small SU}_{{\small 2r}}}$ with respect to the outer-automorphism group $%
\left( \mathbb{Z}_{2}^{{\small outer}}\right) _{Q^{{\small SU}_{{\small 2r}%
}}}$. The quiver series $Q_{\hat{X}\left( Y^{2r,0}\right) }^{{\small SP}_{%
{\small 2r}}}$ is also related to the symplectic CY graphs $\mathcal{G}_{%
\hat{X}\left( Y^{2r,0}\right) }^{{\small SP}_{{\small 2r}}}$ obtained from
the folding of the unitary $\mathcal{G}_{\hat{X}\left( Y^{2r,0}\right) }^{%
{\small SU}_{{\small 2r}}}$ under the outer-automorphism symmetry $H_{\Delta
^{{\small SU}_{{\small 2r}}}}^{{\small outer}}$. The explicit expression of
the generalised Mori-vectors and representative graph $\mathcal{G}_{\hat{X}%
\left( Y^{2r,0}\right) }^{{\small SP}_{{\small 2r}}}$ as well as the
associated quivers have been omitted for the sake of simplifying the
presentation of the underlying idea. \newline
The second extension regards 5D super QFT models, based on conical
Sasaki-Einstein manifolds Y$^{p,q},$ with gauge symmetries beyond the
unitary SU$\left( r+1\right) $ and the symplectic SP$\left( 2r,\mathbb{R}%
\right) $ groups. These gauge symmetries concern the orthogonal SO$\left(
2r\right) $ and SO$\left( 2r+1\right) $ groups; and eventually the three
exceptional Lie groups $\boldsymbol{E}_{6},\boldsymbol{E}_{7}$ and $%
\boldsymbol{E}_{8}$. For 5D super QFT models with SO$\left( 2r\right) $
gauge symmetry embedded in M-theory on $\hat{X}\left( Y^{p,q}\right) $, one
needs engineering toric Calabi-Yau threefolds $\hat{X}^{p,q}\left(
\boldsymbol{D}_{{\small r}}\right) $ with an SO(2r) gauge fiber. This might
be nicely reached by using the technique of the CY graphs $\mathcal{G}_{\hat{%
X}\left( \boldsymbol{D}_{{\small r}}\right) }^{{\small SO}_{{\small 2r}}}$
used in this study although an explicit check is still missing. This series
of $\mathcal{G}_{\hat{X}\left( \boldsymbol{D}_{{\small r}}\right) }^{{\small %
SO}_{{\small 2r}}}$ could be constructed by taking advantage of known
results from the so-called complex $\boldsymbol{D}_{{\small r}}$ surfaces
describing the resolution of ALE space with SO$\left( 2r\right) $
singularity. The family of the CY graphs $\mathcal{G}_{\hat{X}\left(
\boldsymbol{D}_{{\small r}}\right) }^{{\small SO}_{{\small 2r}}}$ might be
also motivated from the correspondence between eq(\ref{3a}) and eq(\ref{3i})
for simply laced case; see also the correspondence between eq(\ref{2i}) and
eq(\ref{2a}) for non simply laced diagrams. If this SO$\left( 2r\right) $
study can be rigourously performed, one can also use outer-automorphisms of $%
\mathcal{G}_{\hat{X}\left( \boldsymbol{D}_{{\small r}}\right) }^{{\small SO}%
_{{\small 2r}}}$, inherited for the Dynkin diagram of so$\left( 2r\right) $
Lie algebra, as well as the outer-automorphisms of the associated BPS $Q_{%
\hat{X}\left( \boldsymbol{D}_{{\small r}}\right) }^{{\small SO}_{{\small 2r}%
}}$ to construct 5D supersymmetric QFT models with SO$\left( 2r-1\right) $
gauge invariance. Progress in these directions will be reported elsewhere.

\section*{Appendices}

In this section, we give three appendices: A, B and C. They collect useful
tools and give some details regarding the study given in this paper. In
appendix A, we recall general aspects of the families of CY3s used in the
geometric engineering of 5D $\mathcal{N}=1$ super QFTs and the 5D $\mathcal{N%
}=1$ super CFTs. We also describe properties of the Coulomb branch of the 5D
SQFTs. In appendix B, we illustrate the derivation of the formula (\ref{dps}%
). In appendix C, we describe through examples the relationship between the
5D Kaluza-Klein BPS quivers and their 4D counterparts.

\subsection*{Appendix A}

We begin by reviewing interesting aspects of M-theory compactified on a
smooth non compact Calabi-Yau threefold $\hat{X}.$ Then, we focus on
illustrating these aspects for the class of CY3s given by $\hat{X}\left(
Y^{p,q}\right) $ used in present study. We also use these aspects to comment
on the properties of the BPS particle and string states of the 5D gauge
theory.

\subsubsection*{Two local CY3 families}

Generally speaking, we distinguish two main families of local Calabi-Yau
threefolds $\hat{X}$ depending on whether they have an elliptic fibration or
not. These two families are used in the compactification of
F-theory/M-theory/ type II strings leading respectively to effective gauge
theories in 6/5/4 space time dimensions. These compactifications have
received lot of interest in recent years in regards with the full
classification of superconformal theories in various dimensions and their
massive deformations. Because of dualities and due to the biggest 6D, the
classification of 6D effective gauge theories has been conjectured to be the
mother of the classifications in the lower dimensional theories. What
concerns us in this appendix is not the study of the classification issue;
but rather give some mathematical tools developed there and which can also
be applied to our study.

$\bullet $ \emph{Family of local CY3s admitting an elliptic fibration}.
\newline
These local Calabi-Yau threefolds\emph{\ }$\hat{X}$ are complex 3D spaces
given by the typical fibration $E\rightarrow B$ with building blocks as: $%
\left( \mathbf{i}\right) $ B a complex 2D base; this is a Kahler surface. $%
\left( \mathbf{ii}\right) $ a complex 1D fiber E given by an elliptic curve.
This genus zero curve is expressed by the Weirstrass equation%
\begin{equation}
E:y^{2}z=x^{3}+fxz^{2}+gz^{3}  \tag{A.1}
\end{equation}%
where $\left( x,y,z\right) $ are homogeneous coordinates of $\mathbb{P}^{2}.$
Moreover, z is a function on the base B and $\left( x,y,f,g\right) $ are
sections $K_{B}^{-2},K_{B}^{-3},K_{B}^{-4},K_{B}^{-6}$ with $K_{B}$ the
canonical divisor class of B. \newline
Depending on the nature of the base, one can preserve either preserve 16
supersymmetric charges for bases $B$ type $\mathbb{T}^{2}\rightarrow \mathbb{%
P}^{1};$ or eight supercharges in the case of bases $B$ like for example $%
\mathbb{P}^{1}\times \mathbb{P}^{1}$ and in general Hirzebruch surfaces $%
\mathbb{F}_{n}$. These elliptically fibered CY3 geometries $\hat{X}\sim
E\times B$ have been used recently in the engineering of superconformal
theories in dimensions bigger than 4D. Regarding the SCFTs in 4D, the
classification has been obtained a decade ago by using type II strings. For
the classification of the 5D SCFTs using M-theory on elliptically fibered
CY3 we refer to \cite{2ab}. The graphs representing these theories are
intimately related with the Dynkin diagrams of affine Kac-Moody Lie algebras.

$\bullet $ \emph{Family of local CY3s not elliptically fibered}. \newline
As examples of local Calabi-Yau threefolds\emph{\ }$\hat{X}$, we cite the
orbifolds of the complex 3- dimension space; i.e $\mathbb{C}^{3}/\Gamma $
with discrete group $\Gamma $ contained in SU$\left( 3\right) $. These
orbifolds include the conical Sasaki-Einstein threefolds $\hat{X}\left(
Y^{p,q}\right) $ we have considered in this paper. The local CY3 geometries
which are\emph{\ }not elliptically fibered are used in the engineering of
massive supersymmetric QFTs. The graphs representing these theories are
related with the Dynkin diagrams of ordinary Lie algebras.\newline
In what follows, we focus on M-theory compactified on $\hat{X}\left(
Y^{p,0}\right) $ considered in this study and on the corresponding U$\left(
1\right) ^{p-1}$ Coulomb branch.

\subsubsection*{M-theory on $\hat{X}\left( Y^{p,0}\right) $}

The local threefolds $\hat{X}\left( Y^{p,0}\right) $ has four non compact
divisors $\left \{ D_{i}\right \} _{1\leq i\leq 4}$ and p-1 compact divisors
$\left \{ E_{a}\right \} _{1\leq a\leq p-1}$. These divisors are not
completely free; they obey some constraint relations; in particular the
Calabi Yau condition of $\hat{X}\left( Y^{p,0}\right) .$ They also obey
\textrm{gluing properties through holomorphic curves. }The CY condition
reads in terms of the divisor classes as in eq(\ref{cyde}). For a generic
positive integer p; it reads as follows \cite{1e}%
\begin{equation}
\sum_{i=1}^{4}D_{i}+\sum_{a=1}^{p-1}E_{a}=0  \tag{A.2}
\end{equation}%
In our study, this condition has been transformed as in eq(4.1); and has
been used to introduce the graphs given in section 4. Notice that the union
of the compact divisors $S=\cup _{a=1}^{p-1}E_{a}$ is important in this
investigation; it is a local surface made of a collection of irreducible
compact holomorphic surfaces $E_{a}.$ The irreducible holomorphic surfaces
intersect each other pairwise transversally; this intersection is important
and will be described below with details. Notice also that the Kahler
parameters of the $E_{a}$'s are identified as the Coulomb branch moduli $%
\phi _{a};$ they appear in the calculation through the linear combination $%
\sum \phi _{a}E_{a}$ which also plays an important role in the construction.
\newline
Regarding the gluing properties of the compact divisors and their
consequences; they need introducing some geometric tools of the CY3. For a
shortness and self contained of the presentation, we restrict to giving only
those main tools that are interesting for this study. However, we take the
occasion to also describe some particular geometric objects that are
relevant for the investigation of the Coulomb branch of the gauge theory.
These geometric objects are introduced through the four following points
(a), (b), (c) and (d).

$\mathbf{a})$ \emph{Gluing the compact divisors}\newline
The compact holomorphic surfaces $\left \{ E_{a}\right \} $ are complex
surfaces in $\hat{X}.$ Neighboring surfaces $E_{a}$ and $E_{b}$ are glued to
each others while satisfying consistency conditions. Before giving these
conditions, recall that in our study, we have solved the CY condition by
thinking of the $E_{a}$'s as given by $\left( \mathbb{F}_{0}\right) _{a}$.
As the holomorphic surface $\mathbb{F}_{0}$ is given by a projective line $%
\mathbb{P}_{f}^{1}$ trivially fibered over a base $\mathbb{P}_{B}^{1}$, then
we have
\begin{equation}
E_{a}=\left( \mathbb{P}_{f}^{1}\right) _{a}\times \left( \mathbb{P}%
_{B}^{1}\right) _{a}  \tag{A.3}
\end{equation}%
Notice that this is a particular solution of the CY on 4-cycles; it has been
motivated by looking for a simple solution to exhibit the CY condition as in
the Figures 7-8-9 of section 4. However, general solutions might be worked
out by using other type of holomorphic compact surfaces like the Hirzebruch
surfaces $\left( \mathbb{F}_{n}\right) _{a}$ of degree n and their blow ups
at generic points. To fix the ideas, we\ focus below on the surfaces $%
\mathbb{F}_{n}$ and on two lattices associated with $\mathbb{F}_{n}$ namely:
$\left( \mathbf{1}\right) $ the lattice $\Lambda _{\QTR{sl}{l}}\left(
\mathbb{F}_{n}\right) $ of complex curves $\QTR{sl}{l}$ in $\mathbb{F}_{n}$;
and $\left( \mathbf{2}\right) $ the Mori cone of curves $\mathfrak{M}_{%
\QTR{sl}{l}}\left( \mathbb{F}_{n}\right) $; this is a particular sublattice
of $\Lambda _{\QTR{sl}{l}}\left( \mathbb{F}_{n}\right) .$ To that purpose,
recall that holomorphic curves $\QTR{sl}{l}$ in the compact surface $\mathbb{%
F}_{n}$ are generated by two basic (irreducible) curves e and f. The base
curve $e$ is the zero section of the fibration; and the f is the fiber $%
\mathbb{P}_{f}^{1}$. The intersection numbers of these generators are given
by%
\begin{equation}
e^{2}=-n\qquad ,\qquad f^{2}=0\qquad ,\qquad e.f=1  \tag{A.4}
\end{equation}%
Before proceeding forward, notice the four following interesting aspects: $%
\left( \mathbf{i}\right) $ The positivity $e.f\geq 0$ captures the
irreducibility property of the generators e and f. In general, a given curve
$\QTR{sl}{l}$ belonging to $\Lambda _{\QTR{sl}{l}}\left( \mathbb{F}%
_{n}\right) $ is said irreducible if we have $\QTR{sl}{l}.e\geq 0$ and $%
\QTR{sl}{l}.f\geq 0$. $\left( \mathbf{ii}\right) $ As far compact
holomorphic curves in $\mathbb{F}_{n}$ are concerned; one distinguishes two
interesting curves that play an important role in the study of $\mathbb{F}%
_{n}$. These are the curve $h=e+nf;$ and the canonical class $K_{\mathbb{F}%
_{n,g}}=-2h+(2g-2+n)f$ where we have moreover figured the genus g. For the
case of $\mathbb{F}_{n}$, we have $K_{\mathbb{F}_{n}}=-2e+\left( n-2\right)
f;$ it reduces for the case $n=0$ to $K_{\mathbb{F}_{0}}=-2e-2f$ with triple
intersection given by%
\begin{equation}
\left( \mathbb{F}_{0}\right) ^{3}=K_{\mathbb{F}_{0}}.K_{\mathbb{F}%
_{0}}=8e.f=8  \tag{A.5}
\end{equation}%
From the above relations, we can perform several computations. For example,
we have%
\begin{equation}
\begin{tabular}{llllllll}
$h^{2}$ & $=+n$ & $\qquad ,\qquad $ & $h.e$ & $=0$ & $\qquad ,\qquad $ & $%
h.f $ & $=1$ \\
$K_{\mathbb{F}_{0}}.h$ & $=-2$ & $\qquad ,\qquad $ & $K_{\mathbb{F}_{0}}.e$
& $=2n-2$ & $\qquad ,\qquad $ & $K_{\mathbb{F}_{0}}.f$ & $=-2$%
\end{tabular}
\tag{A.6}
\end{equation}%
and
\begin{equation}
\left( K_{\mathbb{F}_{n,g}}+\QTR{sl}{l}\right) .\QTR{sl}{l}=2g-2  \tag{A.7}
\end{equation}%
where $g$ is the genus of $\QTR{sl}{l}$. Because the genus $g\geq 0$, the
above quantity is greater than $-2$ due to the constraint $2\left(
g-1\right) \geq -2$. $\left( \mathbf{iii}\right) $ Holomorphic curves in
Mori cone $\mathfrak{M}_{\QTR{sl}{l}}\left( \mathbb{F}_{n}\right) $ of the
surface $\mathbb{F}_{n}$ are given by the linear combination \textsl{l}$%
_{n_{e},n_{f}}=n_{e}e+n_{f}f$ with positive integers $n_{e}$ and $n_{f}$.
These are particular curves of $\Lambda _{\QTR{sl}{l}}\left( \mathbb{F}%
_{n}\right) $ corresponding to $n_{e}$ and $n_{f}$ arbitrary integers$.$
Notice that with this notation, we have $h=$\textsl{l}$_{1,n};$ and the
particular curve \textsl{l}$_{1,1}=e+f$ has a self intersection \textsl{l}$%
_{1,1}^{2}=2-n$. $\left( \mathbf{iv}\right) $ If considering several
surfaces $\mathbb{F}_{n_{a}}$ with a=1,...,p-1; then eq(A.4) extend as
follows $e_{a}^{2}=-n_{a}$ and $f_{a}^{2}=0$ as well as $e_{a}.f_{a}=1.$
Quite similar relationships can be written down for the holomorphic curves
in $\Lambda _{\QTR{sl}{l}}^{a}=\Lambda _{\QTR{sl}{l}}\left( \mathbb{F}%
_{n_{a}}\right) $ and curves in $\mathfrak{M}_{\QTR{sl}{l}}^{a}=\mathfrak{M}%
_{\QTR{sl}{l}}\left( \mathbb{F}_{n_{a}}\right) $. \newline
Returning to the gluing of curves $\QTR{sl}{l}_{a}$ and $\QTR{sl}{l}_{b}$
inside two compact surfaces S$_{a}$ and S$_{b}$; say the divisor E$_{a}$ and
the divisor E$_{b}.$ It is defined by using the following restrictions%
\begin{equation}
\QTR{sl}{l}_{ab}=\left. \QTR{sl}{l}_{a}\right \vert _{E_{b}}\qquad ,\qquad
\QTR{sl}{l}_{ba}=\left. \QTR{sl}{l}_{b}\right \vert _{E_{a}}  \tag{A.8}
\end{equation}%
and imposing some consistency conditions coming from topology and geometry.
The topology requires the two curves $\QTR{sl}{l}_{a}$ and $\QTR{sl}{l}_{b}$
to be identical in the following sense: $\left( \mathbf{\alpha }\right) $ if
the $\QTR{sl}{l}_{a}$ is irreducible; then the $\QTR{sl}{l}_{b}$ must be
also irreducible; and $\left( \mathbf{\beta }\right) $ the genera of the two
curves have to be equal and positive; that is $g\left( \QTR{sl}{l}%
_{a}\right) =g\left( \QTR{sl}{l}_{b}\right) \geq 0$. The geometry requires
moreover the volumes vol$\left( \QTR{sl}{l}_{ab}\right) $ and vol$\left(
\QTR{sl}{l}_{ba}\right) $ to be equal; these volumes are computed by using
the dual Kahler divisor as usual like $-J.\QTR{sl}{l}_{ab}=-J.\QTR{sl}{l}%
_{ba}.$ Under these consistency conditions, the gluing of the two curves is
thought of in terms of the \emph{identification} $\QTR{sl}{l}_{ab}\simeq
\QTR{sl}{l}_{ba}$ together with the CY condition that reads as follows
\begin{equation}
\left( \QTR{sl}{l}_{ab}\right) ^{2}+\left( \QTR{sl}{l}_{ba}\right) ^{2}=2g-2
\tag{A.9}
\end{equation}%
where $g$ is the genus of $\QTR{sl}{l}_{ab}\simeq \QTR{sl}{l}_{ba}$. For the
solution of section 4, the surfaces S$_{a}$ and S$_{b}$ are given by $\left(
\mathbb{F}_{0}\right) _{a}$ and $\left( \mathbb{F}_{0}\right) _{b};$ and the
curves $\QTR{sl}{l}_{ab}$ and $\QTR{sl}{l}_{ba}$ may be taken as $\left(
e_{a}+f_{a}\right) _{b}$ and $\left( e_{b}+f_{b}\right) _{a}.$

$\mathbf{b})$\emph{\ Compact holomorphic curves in} $\hat{X}\left(
Y^{p,0}\right) $\newline
The compact holomorphic curves $C$ in $\hat{X}\left( Y^{p,0}\right) $ are
2-cycles in the local Calabi-Yau threefolds. A subset of these curves is
given by the $e_{a}$'s and the $f_{a}$'s generating the curves in the
divisors $E_{a}$ when realised in terms of $\left( \mathbb{F}_{0}\right)
_{a} $. In general, the compact curves $C$ are given by linear combinations
of generators $C_{\tau }$ of compact holomorphic curves in $\hat{X}\left(
Y^{p,0}\right) ;$ they can be denoted like $C_{\mathbf{n}}$ where $\mathbf{n}
$ is an integer vector. As we have done\ above for the irreducible gauge
divisors $E_{a}=\left( \mathbb{F}_{0}\right) _{a}$, these CY3 holomorphic
curves can be expressed as integer linear combinations like
\begin{equation}
C_{\mathbf{n}}=\sum_{\tau =1}^{d}n_{\tau }C_{\tau }  \tag{A.10}
\end{equation}%
with $n_{\tau }\in \mathbb{Z}$. From this expansion, we learn: $(\mathbf{i})$
the set of compact holomorphic curves in $\hat{X}\left( Y^{p,0}\right) $
form a d-dimensional lattice $\Lambda _{\QTR{sl}{C}}(\hat{X})$ contained in $%
\mathbb{Z}^{d}.$ $(\mathbf{ii})$ In the case where all $n_{\tau }$ integers
are positive ($n_{\tau }\in \mathbb{Z}^{+}$); the corresponding holomorphic
curves belong to Mori cone $\mathfrak{M}_{\QTR{sl}{C}}(\hat{X})$.

$\mathbf{c})$ \emph{Curves intersecting surfaces}\newline
This is an interesting intersection product defined in the CY3. Given the
two following : (i) a holomorphic curve $\QTR{sl}{l}$ belonging to the Mori
cone $\mathfrak{M}(\hat{X})$. (ii) a holomorphic surface $S$ with canonical
class K$_{S}$ sitting in the local Calabi-Yau threefolds $\hat{X}.$ Then,
the intersection between $\QTR{sl}{l}$ and $S$ is given by
\begin{equation}
\QTR{sl}{l}.S=\left. \left( \QTR{sl}{l}.K_{S}\right) \right \vert _{S}
\tag{A.11}
\end{equation}%
For the interesting case where the holomorphic surface $S$ is given by the
compact divisors $E_{a},$ the above intersection reads as $\left. \left(
\QTR{sl}{l}.K_{S}\right) \right \vert _{E_{a}}.$ The value of this
intersection depends on two possibilities: \newline
$\left( \mathbf{\alpha }\right) $ The case where $\QTR{sl}{l}$ lives inside $%
\mathfrak{M}_{\QTR{sl}{a}}(\hat{X});$ then, we have $\QTR{sl}{l}%
.E_{a}=\left. \left( \QTR{sl}{l}.K_{S}\right) \right \vert _{E_{a}}.$\newline
$\left( \mathbf{\beta }\right) $ The case where $\QTR{sl}{l}$ lives inside
another surface; say $S=E_{b};$ then we have
\begin{equation}
\QTR{sl}{l}.E_{a}=\left. \left( \QTR{sl}{l}.\QTR{sl}{l}_{ba}\right) \right
\vert _{E_{b}}  \tag{A.12}
\end{equation}%
where $\QTR{sl}{l}_{ba}$ is the curve participating in the gluing between $%
E_{a}$ and $E_{b}.$ The curve $\QTR{sl}{l}_{ab}$ also sits in $\mathfrak{M}_{%
\QTR{sl}{a}}(\hat{X})$. From these relations, we learn that the
intersections $\QTR{sl}{l}.E_{a}$ can be recovered from the intersection
products on the Mori cones $\mathfrak{M}_{\QTR{sl}{a}}$.

$\mathbf{d})$ \emph{Triple intersections}\newline
The triple intersections $E_{a}.E_{b}.E_{c}$ of the holomorphic surfaces are
numbers that can be expressed as intersection products of gluing curves
inside any of the three surfaces. For that, we use the typical curves $%
\mathcal{L}_{ab}=E_{a}.E_{b};$ these intersection curves appear as
irreducible curves $\QTR{sl}{l}_{ab}$ from the $E_{a}$ side; and as
irreducible curves $\QTR{sl}{l}_{ba}$ from the side of $E_{b}$. The
intersection of $E_{a}$ and $E_{b}$ is obtained as described before; that is
by the identification $\QTR{sl}{l}_{ba}=\QTR{sl}{l}_{ab}$. Similar
identifications hold for the intersections of $E_{b}.E_{c}$ and $%
E_{c}.E_{a}. $ By taking the intersection curve $\QTR{sl}{l}_{\alpha \beta }$
as the diagonal sum of the the generators namely $\QTR{sl}{l}_{\alpha \beta
}=e_{\alpha }+f_{\alpha },$ we obtain
\begin{equation}
\begin{tabular}{lllll}
$E_{a}^{3}$ & $=$ & $K_{\left( \mathbb{F}_{0}\right) _{a}}.K_{\left( \mathbb{%
F}_{0}\right) _{a}}$ & $=+8e_{a}.f_{a}$ & $=+8$ \\
$E_{a}^{2}.E_{b}$ & $=$ & $K_{\left( \mathbb{F}_{0}\right) _{a}}.\QTR{sl}{l}%
_{ab}$ & $=-2e_{a}.f_{a}$ & $=-2$%
\end{tabular}
\tag{A.13}
\end{equation}%
in agreement with eq(\ref{qa}).

\subsubsection*{5D Coulomb branch and BPS states}

To deal with the Coulomb branch of the 5D effective gauge theory and its BPS
states, we need, in addition to the algebraic geometric objects given above,
other basic quantities. One of these quantities concerns the metric $%
ds^{2}=\tau _{ab}d\phi ^{a}d\phi ^{b}$ of the Coulomb branch. It turns out $%
\tau _{ab}$ derives from the effective scalar potential $F\left( \phi
\right) $ of the low energy theory; it reads as follows
\begin{equation}
\tau _{ab}=\frac{\partial ^{2}F\left( \phi \right) }{\partial \phi
^{a}\partial \phi ^{b}}  \tag{A.14}
\end{equation}%
Given $F\left( \phi \right) ,$ one also has two other interesting quantities
associated to it. $\left( \mathbf{i}\right) $ the gradient $\frac{\partial
F\left( \phi \right) }{\partial \phi ^{a}}$ which give the tensions $T_{a}$
of BPS string states. $\left( \mathbf{ii}\right) $ the third derivatives as $%
\partial ^{3}F\left( \phi \right) /\partial \phi ^{a}\partial \phi
^{b}\partial \phi ^{c}$ giving coefficient of the Chern-Simons term $\kappa
_{abc}=kd_{abc}$. The higher derivatives vanish identically because $F\left(
\phi \right) $\ is a cubic function. Recall that the effective potential of
the 5D effective theory is exactly known; it reads as follows%
\begin{equation}
F=\frac{1}{2g_{0}^{2}}h_{ab}\phi ^{a}\phi ^{b}+\frac{\kappa }{6}d_{abc}\phi
^{a}\phi ^{b}\phi ^{c}+\frac{1}{12}\left( \sum_{roots\text{ }\mathbf{\alpha }%
}\left \vert \mathbf{\alpha .\phi }\right \vert ^{3}-\sum_{f}\sum_{\omega
\in \mathbf{R}_{f}}\left \vert \mathbf{\omega .\phi }+m_{f}\right \vert
^{3}\right)  \tag{A.15}
\end{equation}%
This function has the properties: $\left( \mathbf{i}\right) $ It is a cubic
function of the gauge scalar field moduli $\left \{ \phi _{1},...,\phi
_{r}\right \} $ parameterising the Coulomb branch. $\left( \mathbf{ii}%
\right) $ It depends on the mass parameters $m_{f}$ of the 5D effective
theory. $\left( \mathbf{iii}\right) $ It also depends on the roots $\mathbf{%
\alpha }$ of Lie algebra and representations weights $\omega \in \mathbf{R}%
_{f}$ of the underlying gauge symmetry group. From the geometric view, the
5D gauge theory has an interesting description in terms of even p-cycles in
the CY3. These cycles captures information whose some are presented through
the four following poinst (a), (b), (c) and (d).

$\left( a\right) $ \emph{Dual of the Kahler 2-form}\newline
The dual of the Kahler 2-form of $\hat{X}\left( Y^{p,0}\right) $ is a
divisor of the CY3 reading in terms of the generating divisors and the
Coulomb branch moduli as follows%
\begin{equation}
J=\sum_{i=1}^{4}m^{i}D_{i}+\sum_{a=1}^{p-1}\phi ^{a}E_{a}  \tag{A.16}
\label{je}
\end{equation}%
The compact complex surfaces $E_{a}$ are in one to one with the $U\left(
1\right) $ factors of the Coulomb branch of the 5D gauge theory. In the
SU(4) gauge theory studied in the paper, we have three $\phi $'s.

$\left( b\right) $ \emph{Volume of compact even p-cycles in }$\hat{X}\left(
Y^{p,0}\right) $\newline
These compact cycles include: $\left( \mathbf{i}\right) $ the set of compact
2- cycles C belonging H$_{2}(\hat{X})$, $\left( \mathbf{ii}\right) $ the set
of compact 4- cycles S belonging H$_{4}(\hat{X})$; and $\left( \mathbf{iii}%
\right) $ the 6-cycle given by $\hat{X}\left( Y^{p,0}\right) $. \newline
The volume of a compact 2-cycles C is given by the intersection number%
\begin{equation}
Vol\left( C\right) =-J.C  \tag{A.17}
\end{equation}%
For the particular compact holomorphic curves given by the $p-1$ curves
basic curves $C_{a}$ generating the Mori cone of $\hat{X}\left(
Y^{p,0}\right) ;$ we have the elementary volumes $Vol\left( C_{a}\right)
=\upsilon _{a}$. \newline
The volume of a compact 4-cycles S is given by the intersection number%
\begin{equation}
Vol\left( S\right) =\frac{1}{2}J.J.S  \tag{A.18}
\end{equation}%
For the particular compact holomorphic surfaces given by the basic $p-1$
divisors E$_{a};$ we have the elementary volumes $Vol\left( E_{a}\right) =%
\tilde{\upsilon}_{a}$. \newline
Finally, the volume of of $\hat{X}\left( Y^{p,0}\right) ;$ it is given by
the triple intersection number of the divisor J. This is the prepotential of
the low energy 5D theory
\begin{equation}
F\left( \phi \right) =-\frac{1}{3!}J.J.J  \tag{A.19}  \label{j3}
\end{equation}%
Notice that by putting (\ref{j3}) back into $T_{a},\tau _{ab},\kappa _{abc}$
and using (\ref{je}), we end up with the following interpretation in terms
of intersections%
\begin{equation}
T_{a}\sim E_{a}.J.J\qquad ,\qquad \tau _{ab}\sim E_{a}.E_{a}.J\qquad ,\qquad
\kappa _{abc}\sim E_{a}.E_{b}.E_{c}  \tag{A.20}
\end{equation}

$\left( c\right) $ \emph{The BPS states of the 5D theory }\newline
In this effective gauge theory, we distinguish two kinds of BPS states:
\newline
$\left( \mathbf{i}\right) $ Massive particle states $\left( M2/C\right) $
given by M2- branes wrapping the compact holomorphic curves C. The masses of
these particle states are given by $Vol\left( C\right) .$ For the particular
compact curves $C_{a};$ it is associated p-1 electrically charged elementary
BPS particles given by the wrapping M2/C$_{a}.$ The masses of these
particles are given by $\upsilon _{a}.$ \newline
$\left( \mathbf{ii}\right) $ String states M5/S arising from M5- brane
wrapping the compact holomorphic surfaces S. The tensions of these strings
are given by $Vol\left( S\right) .$ For the particular compact surfaces
given by the p-1 divisor E$_{a};$ it is associated p-1 magnetically charged
elementary BPS strings M5/E$_{a}$ with tensions given by $\tilde{\upsilon}%
_{a}.$\newline
Notice that the BPS spectrum of 5D N=1 theories include gauge instantons I
in addition to the electrically charged particles and the magnetically
charged monopole strings. The central charges of these particles are given by%
\begin{equation}
Z_{elc}=\sum_{a=1}^{p-1}n_{a}^{(elc)}\phi _{a}+m_{0}I\qquad ,\qquad
Z_{mag}=\sum_{a=1}^{p-1}n_{a}^{(mag)}\frac{\partial F}{\partial \phi _{a}}
\tag{A.21}
\end{equation}%
where $n_{a}^{(elc)},n_{a}^{(mag)}$ are integers. Notice that not every
choice of these integers corresponds to the central charge of a physical
state whose mass or tension has to be positive. The values of these n's are
obtained using BPS quivers and their mutations. Notice also that by
compactifying the 5D gauge theories on a finite circle; we generate a Kaluza
Klein particle states as described in the core of the paper.

$\left( d\right) $ \emph{Dirac pairing}\newline
The intersection numbers $C_{a}.E_{b}$ of compact curves $C_{a}$ and compact
surfaces $E_{b}$ describe the Dirac pairing between the BPS particles and
the BPS strings.

\subsection*{Appendix B}

\begin{figure}[tbph]
\begin{center}
\includegraphics[width=12cm]{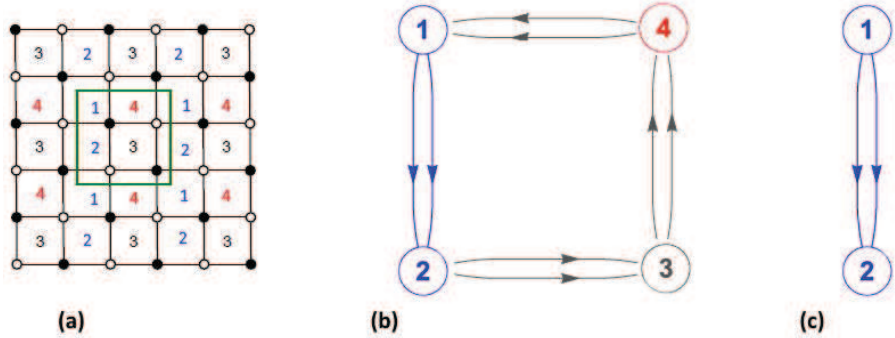}
\end{center}
\par
\vspace{-0.5cm}
\caption{{\protect \small (a) The brane tiling of }$\Delta _{\hat{X}\left(
\mathbb{F}_{0}\right) }^{SU_{2}}${\protect \small . (b) the BPS quiver }$Q_{%
\hat{X}\left( \mathbb{F}_{0}\right) }^{SU_{2}}${\protect \small \ with SU}$%
\left( 2\right) ${\protect \small \ gauge symmetry. (c) the BPS subquiver of
the 4d }$N=2${\protect \small \ pure SU}$\left( 2\right) _{0}${\protect \small %
\ gauge theory.}}
\label{bt23}
\end{figure}

Here, we consider M-theory compactified on $\hat{X}\left( Y^{2,0}\right) $
with SU$\left( 2\right) $ gauge symmetry and look for the derivation of the
quiver dimension $\boldsymbol{d}_{bps}=2\left( p-1\right) +2$ of eq(\ref{dps}%
). Because of the choice p=2, we have $\boldsymbol{d}_{bps}=4$ indicating
that the BPS quiver $Q_{\hat{X}\left( Y^{2,0}\right) }^{SU_{2}}$ has four
nodes as shown by the Figure \textbf{\ref{bt23}}-(b). Recall that the BPS
quiver $Q_{\hat{X}\left( Y^{2,0}\right) }^{SU_{2}}$ is related to the toric
diagram $\Delta _{\hat{X}\left( Y^{2,0}\right) }^{SU_{2}}$ by the so-called
fast inverse algorithm \textrm{\cite{1g,2g,3g}. This algorithm} involves two
main steps summarized as follows:

$\bullet $ \emph{Brane tiling }$BT$\newline
This step maps the toric $\Delta _{\hat{X}\left( Y^{2,0}\right) }^{SU_{2}}$
into a brane tiling in the 2-torus to which we refer to as $BT_{\hat{X}%
\left( Y^{2,0}\right) }$. It \textrm{uses the brane web }$\tilde{\Delta}_{%
\hat{X}\left( Y^{2,0}\right) }^{SU_{2}}$\textrm{\ (the dual of the toric
diagram) }to represent it by the tiling as given by the Figure \textbf{\ref%
{bt23}}-(a).
Recall that the toric graph representing $\Delta _{\hat{X}\left(
Y^{2,0}\right) }^{SU_{2}}$ is a standard diagram; it can be drawn by using
the Table \textbf{\ref{m}} with p=2 and q=0. It has four external points ($%
n_{ext}=4$), describing the four non compact divisors; and one internal
point ($n_{int}=1)$ describing the compact divisor associated with the SU$%
\left( 2\right) $ gauge symmetry. For a short presentation, we have omitted
this graph.

$\bullet $ \emph{The BPS quiver}\newline
The second step maps the brane tiling $BT_{\hat{X}\left( Y^{2,0}\right) }$
into the BPS quiver $Q_{\hat{X}\left( Y^{2,0}\right) }^{SU_{2}}$ as shown by
the picture Figure \textbf{\ref{bt23}}-(b). This mapping is somehow
technical, we propose to illustrate the construction by giving some details.

\subsubsection*{Dimension\textbf{\ }$\boldsymbol{d}_{bps}$ of the quiver $Q_{%
\hat{X}\left( Y^{2,0}\right) }^{SU_{2}}$}

The Figure \textbf{\ref{bt23}}-(a) is a bipartite graph on the 2-torus with
two kinds of nodes white and black. So, half of the nodes are white and the
other half are black. This tiling is characterised by three positive integers%
\textrm{\ }$\left( N_{W},N_{E},N_{F}\right) $\textrm{\ }related amongst
others by the following relation\textrm{\ }%
\begin{equation}
\mathrm{\chi }_{g}=N_{F}-N_{E}+N_{W}  \tag{B.1}  \label{319}
\end{equation}%
where $\mathrm{\chi }_{x g}=2g-2$ is the well known Euler characteristics
relation of discretized real genus- g Rieman surfaces. In this relation: $%
\left( \mathbf{a}\right) $ $N_{W}$ \textrm{i}s the number of nodes\textrm{;}
$\left( \mathbf{b}\right) $\textrm{\ }$N_{E}$\textrm{\ }is the number of
egdes connecting the nodes; and $\left( \mathbf{c}\right) $\ $N_{F}$\ is the
number of faces. This number is precisely the quiver dimension; i.e%
\begin{equation}
N_{F}=\boldsymbol{d}_{bps}  \tag{B.2}
\end{equation}%
Before proceeding, notice that the mapping between a brane tiling $BT_{\hat{X%
}(S)}$\ and a toric diagram $\Delta _{\hat{X}(\boldsymbol{S})}$\ is not
unique. To a given toric diagram one may generally associate several brane
tiling. So the brane tiling is a $1\rightarrow many$. This diversity has an
interpretation in terms of quiver gauge dualities of Seiberg- type. Notice
also that for the 2-torus, we have $g=1$; and then the quiver dimension can
be also expressed as follows
\begin{equation}
N_{F}=N_{E}-N_{W}  \tag{B.3}  \label{fn}
\end{equation}

\subsubsection*{Building the quiver $Q_{\hat{X}\left( Y^{2,0}\right)
}^{SU_{2}}$}

To\textrm{\ build the quiver }$Q_{\hat{X}\left( Y^{2,0}\right) }^{SU_{2}}$
from $BT_{\hat{X}\left( Y^{2,0}\right) }$, we one proceed in steps as
follows:\newline
$\left( \mathbf{i}\right) $ pick up a representative 2-torus unit cell (in
green color in the Figure \textbf{\ref{bt23}}-a).\newline
$\left( \mathbf{ii}\right) $ draw the corresponding BPS quiver given by the
Figure \textbf{\ref{bt23}}-(b) by using the following method.\newline
$\bullet $ To each face F$_{i}$ within the 2-torus unit cell of the
BT-tiling, we associate a quiver- node $\left \{ i\right \} $ in the gauge
quiver $Q_{\hat{X}\left( Y^{2,0}\right) }^{SU_{2}}.$ As there are $N_{F}=4$
faces in the unit cell of $BT_{\hat{X}\left( Y^{2,0}\right) }$, then the $Q_{%
\hat{X}\left( Y^{2,0}\right) }^{SU_{2}}$ has four nodes $\left \{
1;2;3;4\right \} .$ Notice that the number $N_{F}$ can be presented in
different, but equivalent, ways; for instance like $N_{F}=2\left( p-1\right)
+2,$ or%
\begin{equation}
\begin{tabular}{lll}
$N_{F}$ & $=$ & $2r-2+n_{ext}$ \\
& $=$ & $2r+f+1$%
\end{tabular}
\tag{B.4}  \label{nf}
\end{equation}%
where we have set $n_{ext}-2=f+1,$ and where for $SU\left( 2\right) $ the
rank r=1. The number $N_{F}$ is precisely the dimension $\boldsymbol{d}%
_{bps} $\ given by eq(\ref{dps}).\newline
$\bullet $ To each edge $E_{ij}$ of the brane tiling, separating the faces $%
F_{i}$ and $F_{j}$, we associate a quiver- arrow $\left \langle
ij\right
\rangle $ with direction determined by a traffic rule. In this
rule, the circulation goes clockwise around white BT- nodes; and
counter-clockwise around black BT-nodes. In the example $Q_{\hat{X}\left(
Y^{2,0}\right) }^{SU_{2}}$; we have $8$ quiver- arrows organized into four
pairs. The are given by arrows $\left \langle i\left( i+1\right)
\right
\rangle $ with $i=1,2,3,4$ $\func{mod}4$.\newline
$\bullet $ To each BT-node in the brane tiling corresponds a superpotential
monomial. So, the full superpotential $W_{\hat{X}\left( Y^{2,0}\right)
}^{SU_{2}}$ associated with the BPS quiver has four monomials; this is
because $N_{W}=4.$ For simplicity, we omit the explicit expression of $W_{%
\hat{X}\left( Y^{2,0}\right) }^{SU_{2}}$.\newline
In the end of this appendix, notice that the four nodes of the \textrm{%
quiver }$Q_{\hat{X}\left( Y^{2,0}\right) }^{SU_{2}}$ are interpreted in type
IIA string as the elementary BPS particles. The particles sitting at the
nodes $\left \{ 1,2\right \} $ correspond to the electrically $D2/\mathfrak{C%
}_{2}$ and the magnetically $D4/\mathfrak{C}_{4}$ charged BPS particles
where $\mathfrak{C}_{n}$ refers to n-cycles in $\hat{X}\left( Y^{2,0}\right)
$. These two nodes form together a sub-graph of the SU$\left( 2\right) $
gauge quiver $Q_{\hat{X}\left( Y^{2,0}\right) }^{SU_{2}}$; it is given by
the usual Kronecker diagram depicted by the Figure \textbf{\ref{bt23}}-(c).
The node $\left \{ 3\right \} $ is associated with the elementary instanton $%
\mathcal{I};$ and the node $\left \{ 4\right \} $ with the elementary
Kaluza- Klein D0. These two nodes form together the Kronecker sub-diagram $%
\left \{ 3,4\right \} $.

\subsection*{Appendix C}

In this appendix, we describe briefly helpful tools regarding the structure
of BPS quivers in 4D $\mathcal{N}=1$ Kaluza-Klein while focussing on those
relevant aspects for our study. The material given below aims facilitating
the reading of section 3. For a rigorous and abstract formulation of BPS
quivers using amongst others the central charges and the Coulomb branch
moduli, we refer to literature in this matter. For instance, the section 2.3
of \cite{3e} for 4D KK quivers and \cite{1Z}-\cite{3Z}, \cite{0M}, \cite{13h}%
-\cite{17h} for 4D.

\subsubsection*{ADE gauge models}

A short way to introduce the 4D KK BPS quivers is to go through the well
studied BPS quivers $\mathfrak{Q}_{\hat{X}}^{{\small G}}\left( 4d\right) $
of 4D $\mathcal{N}=2$ gauge theories with ADE gauge symmetries. The use of
these 4D quivers may be motivated from various views in particular from the
three following: \newline
$\left( \mathbf{1}\right) $ $\mathfrak{Q}_{\hat{X}}^{{\small G}}\left(
4d\right) \subset \mathfrak{Q}_{\hat{X}}^{{\small G}}\left( 5d\right) .$ The
quivers $\mathfrak{Q}_{\hat{X}}^{{\small G}}\left( 4d\right) $, which are
described below and mention in the text, appear as sub-quivers of the 4D KK\
BPS quivers $Q_{\hat{X}}^{{\small G}}\left( 5d\right) .$\ For example,
compare the two pictures of the Figure \textbf{\ref{QA2}\ }with the Figures
\textbf{5} and \textbf{6 }in the main text. This feature, which implies that
the BPS states of 4D $\mathcal{N}=2$ belong also to 4D KK $\mathcal{N}=1,$
can be explained by the fact that the 4D $\mathcal{N}=2$ theory corresponds
just to the zero mode of 4D Kaluza-Klein$\  \mathcal{N}=1$ theory.\newline
$\left( \mathbf{2}\right) $ \emph{Type II strings on CY3}. The $\mathfrak{Q}%
_{\hat{X}}^{{\small G}}\left( 4d\right) $ quivers deal with 4D $\mathcal{N}%
=2 $ gauge theories with gauge symmetry G. These SQFTs can be remarkably
embedded in type IIA string on CY3s. However, because of the relationship
between type IIA and M-theory, the 4D $\mathcal{N}=2$ theories can be also
embedded in M-theory on CY3$\times \mathbb{S}^{1}$ which is the mother of 4D
$\mathcal{N}=1$ KK theory. \newline
$\left( \mathbf{3}\right) $ \emph{Quiver mutations and duality}. The BPS
quivers have been widely employed in the case of four-dimensional $\mathcal{N%
}=2$ theories. There, several techniques have been developed to handle them.
Some of these techniques like quiver mutation algorithm apply also to $Q_{%
\hat{X}}^{{\small G}}\left( 5d\right) ;$ these mutation have an
interpretation in terms of Seiberg- like duality. For an explicit study, see
\cite{4Z,15h,17h}. \  \  \  \  \newline
In what follows, we focus on the 4D BPS quivers of pure 4D $\mathcal{N}=2$
gauge theories with gauge invariance G; say of type ADE. A general
description of BPS quivers would also involve flavor matter; but for
convenience, we ignore them here. The determination of the full set of BPS
states of the $\mathcal{N}=2$ SQFT is a complicated issue; but nicely
formulated in terms of BPS quivers $\mathfrak{Q}_{\hat{X}}^{{\small ADE}%
}\left( 4d\right) $. So, the BPS quivers encode the relevant data on the BPS
states of the $\mathcal{N}=2$ SQFT. Their properties depend on the
coordinates of the moduli space of the theory. Depending on the gauge
coupling regime, we distinguish two sets $\{ \mathfrak{Q}_{\hat{X}}^{{\small %
ADE}}\left( 4d\right) _{n}\}_{I,II}$ of BPS quivers termed as \ strong and
weak chambers:

\begin{itemize}
\item \emph{Strong chamber }$\{ \mathfrak{Q}_{\hat{X}}^{{\small ADE}}\left(
4d\right) _{n}\}_{str}.$ This is a finite set of BPS quivers describing the
BPS particle states in the strong chamber. For the derivation of the full
list o3f BPS states for ADE Lie algebras; see for instance \cite{15h} and
references therein.

\item \emph{Strong chamber }$\{ \mathfrak{Q}_{\hat{X}}^{{\small ADE}}\left(
4d\right) _{n}\}_{weak}.$ This is an infinite set of BPS quivers describing
the BPS states in the weak chamber. For a description of this set; see for
instance \cite{17h}.
\end{itemize}

The full content of these BPS chambers can be obtained by constructing all
BPS quivers using mutation algorithm (mutation symmetry group). One of these
BPS quivers is given by the so-called primitive BPS quiver denoted below
like $\mathfrak{\mathring{Q}}_{\hat{X}}^{{\small ADE}}$. This is a basic BPS
quiver made of the elementary BPS states. By applying the mutation algorithm
on $\mathfrak{\mathring{Q}}_{\hat{X}}^{{\small ADE}}$, one generates new
quivers made of BPS states given by composites of the elementary ones. By
repeating this operation several times, one can generate all BPS particles
of the theory. For the strong chambers, the mutation group is finite;
however is it infinite for weak chambers. There, one obtains recursive
relations for the EM charges of the BPS states.

\subsubsection*{Primitive quiver}

\begin{figure}[tbph]
\begin{center}
\includegraphics[width=10cm]{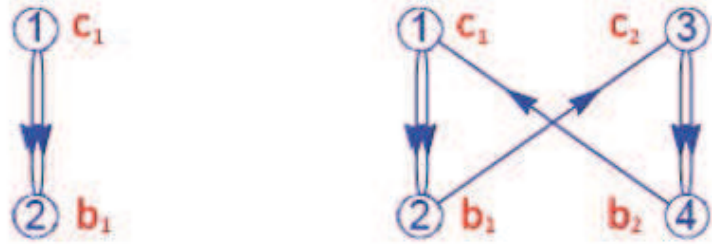}
\end{center}
\par
\vspace{-0.5cm}
\caption{BPS primitive quivers of type ADE in $\mathcal{N}=2$ pure gauge
theories. On left the BPS quiver of pure SU$\left( 2\right) $ gauge model.
It has two nodes and it is often termed as Kronecker quiver. On right the
BPS quiver of SU$\left( 3\right) $ having four nodes. It is made of two
Kronecker quivers with links.}
\label{QA2}
\end{figure}

As far as the primitive quiver of pure gauge theories is concerned, its 2r
BPS states have electric/magnetic (EM) charge vectors given by $\mathbf{b}%
_{1}\mathbf{,...,b}_{r}$ and $\mathbf{c}_{1}\mathbf{,...,c}_{r};$ they
appear in $\mathfrak{\mathring{Q}}_{\hat{X}}^{{\small ADE}}$ as depicted by
the pictures of the Figure \textbf{\ref{QA2} }for SU(2) and SU(3) gauge
groups. The integer r is the rank of the Lie algebra ADE.
The EM charge vectors $\boldsymbol{b}_{i}$ and $\boldsymbol{c}_{i}$ read in
terms of the simple roots $\vec{a}_{1},...,\vec{a}_{r}$ of the simply laced
Lie algebra of the gauge symmetry G as follows%
\begin{equation}
\boldsymbol{b}_{i}=\left(
\begin{array}{c}
\vec{0} \\
\vec{a}_{i}%
\end{array}%
\right) \qquad ,\qquad \boldsymbol{c}_{i}=\left(
\begin{array}{c}
\vec{a}_{i} \\
-\vec{a}_{i}%
\end{array}%
\right)  \tag{C.1}
\end{equation}%
Recall that the simple roots $\vec{a}_{i}$ have the intersection
\begin{equation}
\vec{a}_{i}.\vec{a}_{j}=K_{ij}  \tag{C.2}
\end{equation}%
giving the Cartan matrix of the Lie algebra. Notice that these charge
vectors can be denoted collectively like $\mathbf{b}_{i}=\mathbf{\gamma }%
_{2i}$ and $\mathbf{c}_{i}=\mathbf{\gamma }_{2i-1}$ with $i=1,...,r$. These
EM charge vectors obey the Dirac pairings $\mathbf{\gamma }_{m}\circ \mathbf{%
\gamma }_{n}$ that splits as%
\begin{equation}
\mathbf{\gamma }_{m}\circ \mathbf{\gamma }_{n}=%
\begin{pmatrix}
\boldsymbol{c}_{i}\circ \boldsymbol{c}_{j} & \boldsymbol{c}_{i}\circ
\boldsymbol{b}_{j} \\
\boldsymbol{b}_{i}\circ \boldsymbol{c}_{j} & \boldsymbol{b}_{i}\circ
\boldsymbol{b}_{j}%
\end{pmatrix}%
\equiv \mathcal{A}_{0}^{G}  \tag{C.3}
\end{equation}%
Notice also that for simply laced Lie algebras, the primitive BPS quiver $%
\mathfrak{\mathring{Q}}_{\hat{X}}^{{\small ADE}}$ consists of $2r$ nodes and
$3r-2$ links as described below:

$\mathbf{A})$ \emph{nodes in }$\mathfrak{\mathring{Q}}_{\hat{X}}^{{\small ADE%
}}$\newline
The $2r$ nodes of $\mathfrak{\mathring{Q}}_{\hat{X}}^{{\small ADE}}$ are as
depicted by the pictures of the Figure \textbf{\ref{QA2} }corresponding to%
\textbf{\ }SU$\left( 2\right) $ and SU$\left( 3\right) .$ They\textbf{\ }%
refer to the elementary BPS states with EM charges $\mathbf{b}_{i}$ and $%
\mathbf{c}_{i}$. Notice that in the case of pure gauge theories considered
in this appendix, the $\mathbf{b}_{i}$'s are the charges of r elementary
monopoles $\mathfrak{M}_{i}$ while the $\mathbf{c}_{i}$'s are the charges of
r elementary dyons $\mathfrak{D}_{i}$. So the elementary BPS states of the
theory are given by%
\begin{equation}
\begin{tabular}{lll}
Monopoles & : & $\mathfrak{M}_{1},...,\mathfrak{M}_{r}$ \\
Dyons & : & $\mathfrak{D}_{1},...,\mathfrak{D}_{r}$%
\end{tabular}
\tag{C.4}
\end{equation}%
These BPS states have interpretation in terms of D2- and D4- branes wrapping
2- and 4- cycles in the CY3. In M-theory language, they can be interpreted
in terms of wrapped M2- and M5- branes.

$\mathbf{B})$ \emph{Links between nodes of }$\mathfrak{\mathring{Q}}_{\hat{X}%
}^{{\small ADE}}$\newline
There are $r+\left( 2r-2\right) $ oriented links joining the nodes of\emph{\
}$\mathfrak{\mathring{Q}}_{\hat{X}}^{{\small ADE}}$. These links are of two
types as described below:

\begin{itemize}
\item r vertical links $l_{1},...,l_{r}$ joining the nodes $\left(
\boldsymbol{b}_{i},\boldsymbol{c}_{i}\right) ;$ that is the r pairs $\left(
\boldsymbol{b}_{1},\boldsymbol{c}_{1}\right) ,...,\left( \boldsymbol{b}_{r},%
\boldsymbol{c}_{r}\right) $. They are oriented from the node $\boldsymbol{c}%
_{i}$ to the node $\boldsymbol{b}_{i}.$ These links carry a charge given by
the absolute value of the Dirac pairing $\boldsymbol{b}_{i}\circ \boldsymbol{%
c}_{i}$ which is equal to $2$ due to the relation%
\begin{equation}
\boldsymbol{b}_{i}\circ \boldsymbol{c}_{i}=\vec{a}_{i}.\vec{a}_{i}=2
\tag{C.5}
\end{equation}%
Recall that the Dirac pairing of the electric magnetic charge vectors is
antisymmetric $\mathbf{\gamma }_{m}\circ \mathbf{\gamma }_{n}=-\mathbf{%
\gamma }_{n}\circ \mathbf{\gamma }_{m}$; so we have
\begin{eqnarray}
\boldsymbol{b}_{i}\circ \boldsymbol{c}_{i} &=&-\boldsymbol{c}_{i}\circ
\boldsymbol{b}_{i}=K_{ij}  \TCItag{C.6}  \label{b} \\
\boldsymbol{b}_{i}\circ \boldsymbol{b}_{j} &=&\boldsymbol{c}_{i}\circ
\boldsymbol{c}_{j}=0  \TCItag{C.7}  \label{c}
\end{eqnarray}

\item $2\left( r-1\right) $ oblique links $l_{ij}$ joining two nodes of
different pairs $\left( \boldsymbol{b}_{i},\boldsymbol{c}_{i}\right) $ and $%
\left( \boldsymbol{b}_{j},\boldsymbol{c}_{j}\right) $. This reduced number
of links is due to the constraint eqs(C.6-C.7). So, the intersection matrix $%
\mathcal{A}_{0}^{G}$ describing the primitive quiver $\mathfrak{\mathring{Q}}%
_{\hat{X}}^{{\small ADE}}$ is related to the Cartan matrix of the Lie
algebra as follows%
\begin{equation}
\mathcal{A}_{0}^{G}=%
\begin{pmatrix}
0 & -K \\
K & 0%
\end{pmatrix}
\tag{C.8}  \label{ADE}
\end{equation}
\end{itemize}

This construction extends to the BPS quivers with non simply laced gauge
symmetries; see for instance \cite{0M,13h,14h}.


\begin{thebibliography}{999}
\bibitem{1} G. Zafrir, Brane webs and O5-planes, JHEP03(2016)109,
arXiv:1512.08114 [hep-th].

\bibitem{2} G. Zafrir, Brane webs, 5d gauge theories and 6d $N=(1,0)$
SCFT's, JHEP 12 (2015) 157 [arXiv:1509.02016].

\bibitem{2aa} L. Bhardwaj, Classification of 6d $N=(1,0)$ gauge theories,\
JHEP 11 (2015) 002, arXiv:1502.06594 [hep-th].

\bibitem{2ab} Lakshya Bhardwaj, Patrick Jefferson, Classifying 5d SCFTs via
6d SCFTs: Rank one, JHEP 01 (2020) 153, e-Print: 1809.01650.

\bibitem{2ac} Patrick Jefferson, Sheldon Katz, Hee-Cheol Kim, Cumrun Vafa,
On Geometric Classification of 5d SCFTs, JHEP04(2018)103, arXiv:1801.04036
[hep-th].

\bibitem{3} M.R. Douglas, On D = 5 super Yang-Mills theory and (2; 0)
theory, JHEP 02 (2011) 011, [arXiv:1012.2880].

\bibitem{1C} N. Lambert, C. Papageorgakis and M. Schmidt-Sommerfeld,
M5-Branes, D4-branes and Quantum 5D super-Yang-Mills, JHEP 01 (2011) 083
[arXiv:1012.2882].

\bibitem{1a} Fabio Apruzzi, Ling Lin, Christoph Mayrhofer, Phases of 5d
SCFTs from M-/F-theory on Non-Flat Fibrations, JHEP05(2019)187,
arXiv:1811.12400 [hep-th]

\bibitem{2a} D. Xie and S.-T. Yau, Three dimensional canonical singularity
and five dimensional $N=1$ SCFT, JHEP 06 (2017) 134, arXiv:1704.00799
[hep-th].

\bibitem{3a} S. Alexandrov, S. Banerjee, and P. Longhi, Rigid limit for
hypermultiplets and five-dimensional gauge theories, JHEP 01 (2018) 156,
arXiv:1710.10665 [hep-th]

\bibitem{4a} P. Jefferson, H.-C. Kim, C. Vafa, and G. Zafrir, Towards
Classiffication of 5d SCFTs: Single Gauge Node, arXiv:1705.05836 [hep-th].

\bibitem{5a} Cyril Closset, Michele Del Zotto, Vivek Saxena,
Five-dimensional SCFTs and gauge theory phases: an M-theory/type IIA
perspective, SciPost Phys. 6, 052 (2019), arXiv:1812.10451 [hep-th].

\bibitem{6a} H. Hayashi, S.-S. Kim, K. Lee, M. Taki, and F. Yagi, A new 5d
description of 6d D-type minimal conformal matter,\ JHEP 08 (2015) 097,
arXiv:1505.04439 [hep-th].

\bibitem{7a} J. J. Heckman, D. R. Morrison, and C. Vafa, On the
Classification of 6D SCFTs and Generalized ADE Orbifolds, JHEP 1405 (2014)
028, [arXiv:1312.5746].

\bibitem{7ab} Matteo Sacchi, Orr Sela, Gabi Zafrir, Compactifying 5d
superconformal field theories to 3d, JHEP 09 (2021) 149, arXiv:2105.01497
[hep-th].

\bibitem{7b} L. Bhardwaj, On the classification of 5d SCFTs, JHEP 09 (2020)
007 [arXiv:1909.09635].

\bibitem{11a} Anton Kapustin, On the universality class of little string
theories, Phys.Rev.D 63 (2001) 086005, e-Print: hep-th/9912044.

\bibitem{12a} Chi-Ming Chang, 5d and 6d SCFTs Have No Weak Coupling Limit,
10.1007/JHEP09 (2019) 016, arXiv:1810.04169 [hep-th].

\bibitem{13a} Lakshya Bhardwaj, Patrick Jefferson, Hee-Cheol Kim,
Houri-Christina Tarazi, Cumrun Vafa, Twisted Circle Compactifications of 6d
SCFTs, arXiv:1909.11666 [hep-th].

\bibitem{14a} N. Seiberg, Five-dimensional SUSY field theories, nontrivial
fixed points and string dynamics, Phys. Lett. B 388 (1996) 753
[hep-th/9608111]

\bibitem{1A} O. Aharony and A. Hanany, Branes, superpotentials and
superconformal fixed points, Nucl. Phys. B 504 (1997) 239 [hep-th/9704170].

\bibitem{2A} O. Aharony, A. Hanany and B. Kol, Webs of (p,q) five-branes,
five-dimensional field theories and grid diagrams, JHEP 01 (1998) 002
[hep-th/9710116].

\bibitem{3A} N\textrm{. C. Leung and C. Vafa, Bra}nes and toric geometry,
Adv.Theor.Math.Phys. 2 (1998) 91-118, [hep-th/9711013].

\bibitem{4A} S. Cabrera, A. Hanany and F. Yagi, Tropical Geometry and Five
Dimensional Higgs Branches at Infinite Coupling, JHEP 01 (2019) 068,
[1810.01379].

\bibitem{5A} A. Bourget, J. F. Grimminger, A. Hanany, M. Sperling and Z.
Zhong, Magnetic Quivers from Brane Webs with O5 Planes, JHEP 07 (2020) 204,
[2004.04082].

\bibitem{6A} M. Akhond, F. Carta, S. Dwivedi, H. Hayashi, S.-S. Kim and F.
Yagi, Five-brane webs, Higgs branches and unitary/orthosymplectic magnetic
quivers, JHEP 12 (2020) 164, [2008.01027].

\bibitem{1b} D. R. Morrison and N. Seiberg, Extremal transitions and
five-dimensional supersymmetric field theories, Nucl. Phys. B483 (1997)
229-247,arXiv: hep-th/9609070 [hep-th].

\bibitem{2b} K. A. Intriligator, D. R. Morrison, and N. Seiberg,
Five-dimensional supersymmetric gauge theories and degenerations of
Calabi-Yau spaces, Nucl. Phys. B497 (1997) 56-100, arXiv:hep-th/9702198
[hep-th].

\bibitem{3b} S. H. Katz, A. Klemm, and C. Vafa, Geometric engineering of
quantum field theories," Nucl. Phys. B497 (1997) 173-195,
arXiv:hep-th/9609239 [hep-th].

\bibitem{4b} S. Katz, P. Mayr, and C. Vafa, Mirror symmetry and exact
solution of 4-D N=2 gauge theories: 1., Adv.Theor.Math.Phys. 1 (1998)
53-114, arXiv:hep-th/9706110 [hep-th].

\bibitem{5b} T. J. Hollowood, A. Iqbal, and C. Vafa, Matrix models,
geometric engineering and elliptic genera, JHEP 03 (2008) 069,
arXiv:hep-th/0310272 [hep-th].

\bibitem{Nek1} S.-S. Kim and F. Yagi, Topological vertex formalism with
O5-plane, Phys. Rev. D97 (2018) 026011, [arXiv:1709.01928].

\bibitem{Nek2} M. Aganagic, A. Klemm, M. Marino, and C. Vafa, The
Topological vertex, Commun.Math.Phys. 254 (2005) 425\{478, [hep-th/0305132].

\bibitem{Nek3} A. Iqbal, C. Kozcaz, and C. Vafa, The Refined topological
vertex, JHEP 0910 (2009) 069, [hep-th/0701156].

\bibitem{Nek4} H. Awata and H. Kanno, Refined BPS state counting from
Nekrasov's formula and Macdonald functions, Int.J.Mod.Phys. A24 (2009)
2253-2306, [arXiv:0805.0191].

\bibitem{Nek5} Hirotaka Hayashi, Sung-Soo Kim, Kimyeong Lee, Futoshi Yagi,
5-brane webs for 5d $N=1$ $G_{2}$ gauge theories, JHEP03(2018)125,
arXiv:1801.03916 [hep-th].

\bibitem{Nek6} Lalla Btissam Drissi, Houda Jehjouh, El Hassan Saidi,
Refining the Shifted Topological Vertex, J.Math.Phys.50:013509,2009,
arXiv:0812.0513 [hep-th].

\bibitem{2C} O. DeWolfe, A. Hanany, A. Iqbal and E. Katz, Five-branes,
seven-branes and five-dimensional En field theories, JHEP 03 (1999) 006
[hep-th/9902179].

\bibitem{3C} M.R. Gaberdiel and B. Zwiebach, Exceptional groups from open
strings, Nucl. Phys. B 518 (1998) 151 [hep-th/9709013].

\bibitem{4C} D. Gaiotto and H.-C. Kim, Duality walls and defects in 5d N = 1
theories, JHEP 01 (2017) 019 [arXiv:1506.03871].

\bibitem{5C} H. Hayashi, S.-S. Kim, K. Lee and F. Yagi, 6d SCFTs, 5d
Dualities and Tao Web Diagrams, JHEP 05 (2019) 203 [arXiv:1509.03300].

\bibitem{SP1} \textrm{Hirotaka Hayashi, Rui-Dong Zhu, More on topological
vertex formalism for 5-brane webs with O5-plane, JHEP04(2021)292,
arXiv:2012.13303.}

\bibitem{SP2} H. Hayashi, S.-S. Kim, K. Lee, and F. Yagi, Discrete theta
angle from an O5-plane, JHEP 1711 (2017) 041, arXiv:1707.07181.

\bibitem{V2} F. Benini, S. Benvenuti, and Y. Tachikawa, Webs of five-branes
and N=2 superconformal field theories, JHEP 0909 (2009) 052,
[arXiv:0906.0359].

\bibitem{1B} I. Brunner and A. Karch, Branes and six-dimensional fixed
points, Phys. Lett. B409 (1997) 109-116, [hep-th/9705022].

\bibitem{2B} O. Bergman and G. Zafrir, 5d fixed points from brane webs and
O7-planes, arXiv:1507.03860.

\bibitem{3B} H. Hayashi, S.-S. Kim, K. Lee, M. Taki, and F. Yagi, More on 5d
descriptions of 6d SCFTs, JHEP 10 (2016) 126, [arXiv:1512.08239].

\bibitem{4B} \textrm{Xiaobin Li, Futoshi Yagi,} \textrm{Thermodynamic limit
of Nekrasov partition function for 5-brane web with O5-plane,
arXiv:2102.09482 [hep-th].}

\bibitem{5B} \textrm{Hirotaka Hayashi, Sung-Soo Kim, Kimyeong Lee, Futoshi
Yagi, Rank-3 antisymmetric matter on 5-brane webs, JHEP05(2019)133,
arXiv:1902.04754} [hep-th].

\bibitem{6B} Mohammad Akhond, Federico Carta, Siddharth Dwivedi, Hirotaka
Hayashi, Sung-Soo Kime, Futoshi Yagi, Five-brane webs, Higgs branches and
unitary/orthosymplectic magnetic quivers, JHEP12(2020)164, arXiv:2008.01027
[hep-th].

\bibitem{9b} M. R. Douglas, S. H. Katz, and C. Vafa, Small Instantons, Del
Pezzo Surfaces and Type I' Theory, Nucl. Phys. B497 (1997) 155-172,
arXiv:hep-th/9609071

\bibitem{10b} C. P. Herzog, Exceptional collections and del Pezzo gauge
theories, JHEP 04 (2004) 069, arXiv:hep-th/0310262.

\bibitem{11b} B. Feng, S. Franco, A. Hanany, and Y.-H. He, UnHiggsing the
del Pezzo, JHEP 08 (2003) 058, arXiv:hep-th/0209228 [hep-th].

\bibitem{12b} S. Franco and A. Hanany, Geometric dualities in 4-D field
theories and their 5-D interpretation, JHEP 04 (2003) 043,
arXiv:hep-th/0207006 [hep-th]

\bibitem{1c} \textrm{J. P. Gauntlett}, D. Martelli, J. Sparks, and D.
Waldram, Sasaki-Einstein metrics on $S^{2}\times S^{3}$, Adv. Theor. Math.
Phys. 8 no. 4, (2004) 711-734, arXiv:hep-th/0403002.

\bibitem{2c} \textrm{D. Martelli and J. Sparks}, Toric Sasaki-Einstein
metrics on $S^{2}\times S^{3}$, Phys. Lett. B621 (2005) 208-212, arXiv:
hep-th/0505027 [hep-th].

\bibitem{3c} \textrm{S. Benvenuti, }S. Franco, A. Hanany, D. Martelli, and
J. Sparks, An Infinite family of superconformal quiver gauge theories with
Sasaki-Einstein duals, JHEP 06 (2005) 064, arXiv:hep-th/0411264 [hep-th].

\bibitem{4c} J. P. Gauntlett,\textrm{\ }D. Martelli, J. Sparks, S.-T. Yau,
Obstructions to the existence of Sasaki-Einstein metrics, Commun. Math.
Phys. 273 (2007), 803--827.

\bibitem{5c} A. Futaki, H. Ono, G. Wang, Transverse Kahler geometry of
Sasaki manifolds and toric Sasaki-Einstein manifolds, J. Diff. Geom. 83
(2009), 585--636.

\bibitem{1d} \textrm{Jiakang Bao, Grace Beaney Colverd, Yang-Hui He},
\textrm{Quiver gauge theories: beyond reflexivity, JHEP06(2020)161,
arXiv:2004.05295 [hep-th].}

\bibitem{2d} Olaf Lechtenfeld, Alexander D. Popov, Marcus Sperling, Richard
J. Szabo, Sasakian quiver gauge theories and instantons on cones over lens
5-spaces, 10.1016/j.nuclphysb.2015.09.001, arXiv:1506.02786 [hep-th].

\bibitem{3d} D. Martelli and J. Sparks, Toric geometry, Sasaki-Einstein
manifolds and a new infinite class of AdS/CFT duals, Commun. Math. Phys. 262
(2006) 51--89, [hep-10201].

\bibitem{4d} A. Butti and A. Zaffaroni, From toric geometry to quiver gauge
theory: The equivalence of a-maximization and Z-minimization, Fortsch. Phys.
54 (2006) 309--316, hep-th/0512240.

\bibitem{1e} C. Closset, M. Del Zotto, and V. Saxena, Five-dimensional SCFTs
and gauge theory phases: an M-theory/type IIA perspective,\ SciPost Phys. 6
no. 5, (2019) 052, arXiv:1812.10451 [hep-th].

\bibitem{2e} Vivek Saxena, Rank-two 5d SCFTs from M-theory at isolated toric
singularities: a systematic study, J. High Energ. Phys. 2020, 198 (2020),
arXiv:1911.09574 [hep-th].

\bibitem{3e} Cyril Closset, Michele Del Zotto, On 5d SCFTs and their BPS
quivers. Part I: B-branes and brane tilings, arXiv:1912.13502 [hep-th].

\bibitem{4e} Cyril Closset, Sakura Schafer-Nameki, Yi-Nan Wang, Coulomb and
Higgs Branches from Canonical Singularities: Part 0, JHEP02(2021)003,
arXiv:2007.15600 [hep-th].

\bibitem{5e} Cyril Closset, Simone Giacomelli, Sakura Schafer-Nameki, Yi-Nan
Wang, 5d and 4d SCFTs: Canonical Singularities, Trinions and S-Dualities,
JHEP05(2021)274, arXiv:2012.12827 [hep-th].

\bibitem{13h} R.Ahl Laamara, O.Mellal, E.H.Saidi, BPS $\mathcal{N}=2$
spectra of SO7 and SP4 models, Nucl. Phys B914, 2017, 642-696

\bibitem{14h} R.Ahl Laamara, O.Mellal, E.H.Saidi, BPS states in
supersymmetric G2 and F4 models, Nucl. Phys B920, 2017, 157-191.

\bibitem{15h} Murad Alim, Sergio Cecotti, Clay Cordova, Sam Espahbodi,
Ashwin Rastogi, Cumrun Vafa, N=2 Quantum Field Theories and Their BPS
Quivers, arXiv:1112.3984.

\bibitem{16h} E.H Saidi, Weak Coupling Chambers in N=2 BPS Quiver Theory,
Nuclear Physics B 864 (2012), arXiv:1208.2887.

\bibitem{17h} E.H Saidi, Mutations Symmetries in BPS quiver theory: Building
the BPS Spectra, arXiv:1204.0395, Journal of High Energy Physics, (2012),
Number 8, 18.

\bibitem{1h} A. Hanany and K. D. Kennaway, Dimer models and toric diagrams,
arXiv:hep-th/0503149 [hep-th].

\bibitem{2h} S. Franco, A. Hanany, K. D. Kennaway, D. Vegh, and B. Wecht,
Brane dimers and quiver gauge theories, JHEP 01 (2006) 096,
arXiv:hep-th/0504110 [hep-th]

\bibitem{3h} B. Feng, Y.-H. He, K. D. Kennaway, and C. Vafa, Dimer models
from mirror symmetry and quivering amoebae, Adv. Theor. Math. Phys. 12 no.
3, (2008) 489-545, arXiv:hep-th/0511287 [hep-th].

\bibitem{4h} Y. Imamura, Anomaly cancellations in brane tilings, JHEP 06
(2006) 011, arXiv:hep-th/0605097v2.

\bibitem{5h} A. Hanany and R.-K. Seong, Brane Tilings and Refexive Polygons,
Fortsch. Phys.60 (2012) 695-803, arXiv:1201.2614 [hep-th].

\bibitem{6h} S. Franco, Y.-H. He, C. Sun, and Y. Xiao, A Comprehensive
Survey of Brane Tilings, Int. J. Mod. Phys. A32 no. 23n24, (2017) 1750142,
arXiv:1702.03958 [hep-th]

\bibitem{7h} A. Hanany, C. P. Herzog, and D. Vegh, Brane tilings and
exceptional collections, JHEP 0607 (2006) 001, arXiv:hep-th/0602041 [hep-th].

\bibitem{8h} M. Bender and S. Mozgovoy, Crepant resolutions and brane
tilings II: Tilting bundles, ArXiv e-prints (Sept., 2009) , arXiv:0909.2013
[math.AG].

\bibitem{9h} \textrm{A. Hanany, P. Kazakopoulos, and B. Wecht, A New
infinite class of quiver gauge theories, JHEP 08 (2005) 054,
arXiv:hep-th/0503177 [hep-th].}

\bibitem{10h} M. Cvetic, H. Lu, D. N. Page, and C. N. Pope, New
Einstein-Sasaki spaces in five and higher dimensions, Phys. Rev. Lett. 95
(2005) 071101, hep-th/0504225.

\bibitem{18h} J. Qiu and M. Zabzine, Review of localization for 5d
supersymmetric gauge theories, J. Phys. A50 no. 44, (2017) 443014,
arXiv:1608.02966 [hep-th].

\bibitem{19h} James Sparks, New Results in Sasaki-Einstein Geometry,
Riemannian Topology and Geometric Structures on Manifolds (Progress in
Mathematics), Birkhauser (Nov 2008), arXiv:math/0701518 [math.DG].

\bibitem{ds} L.B Drissi, E.H Saidi; Domain walls in topological tri-hinge
matter, Eur. Phys. J. Plus (2021) 136: 68

\bibitem{5f} M. R. Douglas and G. W. Moore, D-branes, quivers, and ALE
instantons, arXiv:hep-th/9603167.

\bibitem{50f} M. Ait Ben Haddou, A. Belhaj, E. H. Saidi, Geometric
Engineering of N=2 CFT\_ \{4\}s based on Indefinite Singularities:
Hyperbolic Case, Nucl.Phys. B674 (2003) 593-614, arXiv:hep-th/0307244.

\bibitem{M} L.Bonora, R.Savelli, Non simply laced Liealgebras via Ftheory
strings, JHEP 1011 (2010) 025, 1007.4668.

\bibitem{0M} Sergio Cecotti, Michele Del Zotto, 4d N = 2 Gauge Theories and
Quivers: the Non-Simply Laced Case. JHEP 10 (2012) 190, e-Print: 1207.7205
[hep-th].

\bibitem{hyp} El Hassan Saidi, Hyperbolic Invariance in Type II
Superstrings, Talk given at IPM String School and Workshop, ISS2005, January
5-14, 2005, Qeshm Island, IRAN, arXiv:hep-th/0502176.

\bibitem{1g} Masahito Yamazaki, Brane Tilings and Their Applications,
Fortsch.Phys.56:555-686,2008, arXiv:0803.4474 [hep-th].

\bibitem{2g} S. Franco, A. Hanany, D. Martelli, J. Sparks, D. Vegh, B.
Wecht, Gauge theories from toric geometry and brane tilings, JHEP 01 (2006)
128, hep-th/0505211.

\bibitem{3g} A. Butti, D. Forcella, and A. Zaffaroni, The dual
superconformal theory for L(p,q,r) manifolds, JHEP 09 (2005) 018,
hep-th/0505220.

\bibitem{6b} E. Witten, Phase transitions in M theory and F theory,\ Nucl.
Phys. B471 (1996) 195--216, arXiv:hep-th/9603150 [hep-th].

\bibitem{3f} R. Ahl Laamara, M. Ait Ben Haddou, A Belhaj, L. B Drissi, E. H
Saidi, RG Cascades in Hyperbolic Quiver Gauge Theories, Nucl.Phys. B702
(2004) 163-188, arXiv:hep-th/0405222 [hep-th].

\bibitem{4f} M. Ait Ben Haddou, A. Belhaj, E. H. Saidi, Classification of
N=2 supersymmetric CFT$_{4}$s: Indefinite Series, J.Phys. A38 (2005)
1793-1806, arXiv:hep-th/0308005.

\bibitem{6f} Malika Ait Benhaddou, El Hassan Saidi, Explicit Analysis of
Kahler Deformations in 4D N=1 Supersymmetric Quiver Theories, Physics
Letters B575(2003)100-110 [hep-th].

\bibitem{1Z} Alim, M., Cecotti, S., Cordova, C., Espahbodi, S., Rastogi, A.,
Vafa, C.: BPS quivers and spectra of complete N=2 quantum field theories.
Commun. Math. Phys. 323, 1185 (2013). arXiv:1109.4941

\bibitem{2Z} Cecotti, S., Del Zotto, M.: Y systems, Q systems, and 4D N = 2
supersymmetric QFT. J. Phys. A 47, 474001 (2014). arXiv:1403.7613.

\bibitem{3Z} Alim, M., Cecotti, S., Cordova, C., Espahbodi, S., Rastogi, A.,
Vafa, C. N = 2 quantum field theories and their BPS quivers. Adv. Theor.
Math. Phys. 18, 27 (2014). arXiv:1112.3984

\bibitem{4Z} G. Bonelli, F. Del Monte and A. Tanzini, BPS Quivers of
Five-Dimensional SCFTs, Topological Strings and q-Painlev\'{e} Equations,
Ann. Henri Poincar\'{e} (2021).{}
\end{thebibliography}
\end{document}